\documentclass[12pt]{article}
\usepackage{graphicx,amssymb,amsmath,empheq}
\usepackage{authblk}
\usepackage{amsthm}
\usepackage{empheq}
\usepackage{subfigure}
\usepackage{braket}
\usepackage{siunitx}
\usepackage{amsmath,amssymb,amsfonts,stmaryrd}
\usepackage{hyperref}
\hypersetup{colorlinks = true, linkcolor=black, citecolor=black, urlcolor=blue}
\usepackage{url}
\usepackage{dsfont}
\usepackage{comment}
\usepackage{svg}
\usepackage{graphicx}

\theoremstyle{plain}

\usepackage{tikz}
\usepackage{tikz-cd}
\usetikzlibrary{arrows}
\usetikzlibrary{intersections}
\usetikzlibrary{shapes.geometric}
\usetikzlibrary{decorations.pathmorphing, patterns,shapes}
\usetikzlibrary{decorations.markings}

\tikzset{
  mid arrow/.style={postaction={decorate,decoration={
        markings,
        mark=at position .575 with {\arrow[#1]{stealth}}
      }}},
  near arrow/.style={postaction={decorate,decoration={
        markings,
        mark=at position .275 with {\arrow[#1]{stealth}}
      }}},
   far arrow/.style={postaction={decorate,decoration={
        markings,
        mark=at position .800 with {\arrow[#1]{stealth}}
      }}},
}

\usepackage[nosort]{cite}

\newtheorem{definition}{Definition}

\renewcommand{\bar}{\overline}
\renewcommand{\tilde}{\widetilde}

\renewcommand{\leq}{\leqslant}
\renewcommand{\geq}{\geqslant}

\newcommand{\Tr}{\operatorname{Tr}}

\newcommand{\dkap}{\delta\kern-1.25pt\varkappa}

\newcommand{\ZZ}{\mathbb{Z}}

\newcommand{\calE}{\mathcal{E}}

\newcommand{\calM}{\mathcal{M}}

\newcommand{\calR}{\mathcal{R}}

\newcommand{\sn}{\operatorname{sn}}

\newcommand{\Wg}{\operatorname{Wg}}

\newcommand{\st}{\operatorname{st}}

\newcommand{\boldu}{\boldsymbol{u}}
\newcommand{\boldv}{\boldsymbol{v}}

\newcommand{\boldx}{\boldsymbol{x}}
\newcommand{\boldy}{\boldsymbol{y}}

\newcommand{\bolds}{\boldsymbol{s}}
\newcommand{\boldone}{\boldsymbol{1}}

\newcommand{\Ens}{\mathbb{E}}

\makeatletter

\newcommand*{\wideboxed}[1]{\setlength{\fboxsep}{1ex}%
  \fbox{\m@th$\displaystyle#1$}}
\makeatother

\title{
Quantum magic dynamics in random circuits
}

\author[1,2]{Yuzhen Zhang}
\author[1]{Yingfei Gu}

\affil[1]{\normalsize\it Institute for Advanced Study, Tsinghua University, Beijing, 100084, China}
\affil[2]{\normalsize\it Department of Physics, University of California, Santa Barbara, California 93106, USA}

\date{October, 28, 2024}

\begin{document}
\maketitle
\begin{abstract}
Magic refers to the degree of ``quantumness'' in a system that cannot be fully described by stabilizer states and Clifford operations alone. In quantum computing, stabilizer states and Clifford operations can be efficiently simulated on a classical computer, even though they may appear complicated from the perspective of entanglement. In this sense, magic is a crucial resource for unlocking the unique computational power of quantum computers to address problems that are classically intractable. 
Magic can be quantified by measures such as Wigner negativity and mana that satisfy fundamental properties such as monotonicity under Clifford operations.  
In this paper,
we generalize the statistical mechanical mapping methods of random circuits to the calculation of Rényi Wigner negativity and mana. Based on this, we find: (1) a precise formula describing the competition between magic and entanglement in many-body states prepared under Haar random circuits; (2) 
a formula describing the the spreading and scrambling of magic in states evolved under random Clifford circuits; (3) a quantitative description of 
magic ``squeezing'' and ``teleportation'' under measurements. 
Finally, we comment on the relation between coherent information and magic.

\end{abstract}

\tableofcontents

\section{Introduction}

Quantum information science has greatly influenced research in condensed matter physics by providing new perspectives on understanding complex phases of matter, such as quantum entanglement~\cite{RevModPhys.80.517,RevModPhys.81.865}. 
However, entanglement alone is not enough to characterize the ``quantumness'' of a system. For example,
the Gottesman-Knill theorem~\cite{Gottesman:1998hu} states that
the stabilizer states (Pauli eigenstates), which can be highly entangled, are nevertheless classically simulatable. 
Therefore, to better understand the quantum nature of a many-body state, it is crucial to consider more refined quantities that can capture the hardness of the classical simulation. 

This quest has led to a concept known as quantum ``magic''.
The term emerged from the study of stabilizer formalism~\cite{Calderbank:1996hm,Gottesman:1997zz} in fault-tolerant quantum computation, where a fundamental resource - magic states~\cite{PhysRevA.71.022316,knill05}, is required to facilitate universal quantum computing. 
Therefore, understanding how to efficiently generate and manage magic is also crucial for practical applications in fault-tolerant quantum computing. 
Given the importance of the subject, several theoretical frameworks, such as resource theory~\cite{Veitch_2012,Emerson:2013zse,PhysRevLett.118.090501,PhysRevA.97.062332,seddon2019quantifying,
Wang_2019,Liu:2020yso}, have been developed to quantify magic in quantum systems. Recently, the dynamics of magic receives significant attention~\cite{Rattacaso:2023kzm,Lopez:2024jjq,Passarelli:2024lpm,Leone:2023uqw,Niroula:2023meg,Tarabunga:2024din,Bejan:2023zqm,Fux:2023brx,Wang:2023uog,Zhou_2020,Turkeshi:2024pnj,Chen:2022yza}. 


However, unlike quantum entanglement, ``quantum magic" currently lacks a clear physical interpretation. Measures of quantum entanglement, such as entanglement entropy, have a profound connection to thermodynamic entropy. Growing evidence suggests that thermodynamic entropy can be understood as emerging from quantum entanglement, especially in isolated quantum systems~\cite{deutsch1991quantum,srednicki1994chaos} and black holes~\cite{bekenstein1973black,hawking1975particle}. In the context of holography, entanglement entropy in the boundary conformal field theory is related to geometric properties in the bulk Anti-de Sitter space~\cite{ryu2006holographic}. In contrast, parallel frameworks and measures for characterizing quantum magic are still underdeveloped, making its understanding more elusive.

In this paper, we employ a statistical mechanical mapping to explore the properties of Wigner negativity and Mana~\cite{Veitch_2012,PhysRevLett.115.070501} in random circuits, including Haar random and random Clifford circuits. Wigner negativity and Mana are key measures of quantum magic, satisfying fundamental properties such as monotonicity under free operations. Our results uncover facets of many-body quantum behavior beyond entanglement in random unitary circuits, which have been instrumental in advancing our understanding of many-body quantum dynamics~\cite{Fisher:2022qey}. This mapping provides valuable insights into the nature of quantum magic. Furthermore, adopting a many-body approach to random circuits offers promising avenues for addressing computational challenges in the quantum computing community.

The remainder of the paper is organized as follows: In Section~\ref{sec: pre}, we establish the notation and introduce the magic measures—Wigner negativity and Mana—used throughout the main text. In Section~\ref{sec:RHC}, we examine a simple setup in which a Haar random unitary circuit generates and spreads magic, revealing an intriguing competition between magic and entanglement entropy. Section~\ref{sec: RCC} explores a more sophisticated scenario involving random Clifford circuits, where magic is spread and scrambled. We also investigate the effects of measurements within Clifford circuits, which can concentrate and teleport magic. We further connect magic with coherent information. In Section~\ref{sec: summary}, we summarize our findings and discuss open questions for future research. Appendix~\ref{appendix: SRE} presents a statistical mechanical mapping for a computable measure of magic—stabilizer Rényi entropy—under Haar random circuits and random Clifford circuits.\footnote{ Notably, this measure is not universally monotonic~\cite{Haug:2023hcs}, as \cite{Haug:2023hcs} identified counterexamples demonstrating non-monotonicity for stabilizer protocols with Rényi index 
$0\leq n<2$; while \cite{Leone:2024lfr} established monotonicity for 
$n \geq 2$. \cite{Wang:2023uog} further extended the definition of stabilizer Rényi entropy to qudits.
}

\section{Wigner negativity and mana}
\label{sec: pre}

Wigner negativity is a tractable magic measure, satisfying monotonicity under free operations such as Clifford unitaries, partial traces and computational basis measurements~\cite{Emerson:2013zse}. It is generally defined for qudit with odd prime dimensions. 

Let us consider an odd prime $q$ and the (discrete) position basis $|n\rangle$ labeled by elements in a finite field $\mathbb{F}_q$, i.e. $|n\rangle \equiv |n+q\rangle$ with $n\in \ZZ$. They span a Hilbert space of dimension $q$. The corresponding 
Pauli X and Z act as 
\begin{equation}
    X\ket{n}=\ket{n+1} ,\qquad
    Z\ket{n}=w^n\ket{n},\quad \text{where} \quad w=e^{2\pi i/q}.
\end{equation}
They satisfy $X^q=Z^q=1$ and $ZX=wXZ$. In the dual (discrete) momentum basis
$\ket{\tilde{m}}=\frac{1}{\sqrt{q}}\sum_{n=0}^{q-1} w^{mn}\ket{n}$, with $m\in \ZZ$ and $|\tilde{m}\rangle \equiv |\tilde{m+q}\rangle$. 
The Pauli $Z$ acts as 
$Z\ket{\tilde{m}}=\ket{\tilde{m+1}}$.  
In other words, $Z$ and $X$ can be understood as the discrete position and momentum operator for the qudit space in question. With this analog, we can label the discrete phase space by integer pairs $(m,n)$ mod $q$ and define the corresponding Wigner function for a density matrix $\rho$ 
\begin{equation}
    W_\rho(m,n):=\frac{1}{q}\sum_{y=0}^{q-1}w^{my}\Big\langle n-2^{-1} y\Big|\rho\Big|n+2^{-1} y\Big\rangle.
    \label{def}
\end{equation}
Note the $2^{-1}$ factor in the bra (ket) should be understood in the finite field $\mathbb{F}_q$, i.e. $2^{-1}=(q+1)/2$ for odd prime $q$. 
This definition parallels the conventional Wigner function in continuous variable systems, i.e. $W_\rho(p,x)=\int \frac{dy}{2\pi} \ e^{ipy}\big\langle x-\frac{y}{2}\big|\rho\big| x+\frac{y}{2}\big\rangle$.

For our purpose, it is convenient  to rewrite the definition \eqref{def} in terms of the phase point operator 
\begin{equation}
A_{(m,n)}=\sum_{y=0}^{q-1}w^{my}\Big|n+2^{-1} y\Big\rangle\Big\langle n-2^{-1} y\Big| \quad \text{and} \quad
    W_\rho(m,n)=\frac{1}{q}\Tr[A_{(m,n)}\rho]. 
\end{equation}
The phase point operator $A_{\boldu}$ (abbreviation $\boldu=(m,n)$ is used for simplicity) can be expressed as the parity operator conjugated by a displacement operator
\begin{equation}
    A_{\boldu}=T_{\boldu}A_0T_{\boldu}^\dagger,\qquad A_0=\sum_{j=0}^{q-1}\ket{j}\bra{-j}, \label{eq:phase_point_operators}
\end{equation}
where $A_0$ is the parity operator in the qudit space.\footnote{An equivalent definition is $A_{\boldu}=q^{-1}\sum_{\boldv}w^{[\boldu,\boldv]}T_{\boldv}$. To show that they are equivalent: $A_{(m,n)}=q^{-1}\sum_{k,l}w^{ml-nk-kl/2}Z^kX^l=q^{-1}\sum_{k,l,x}w^{ml+k(x-nl/2)}\ket{x+l}\bra{x}=\sum_{l}w^{ml}\ket{n+l/2}\bra{n-l/2}$.} The displacement operators (also known as the Heisenberg-Weyl operators) $ T_{(m,n)}=w^{- 2^{-1}mn}Z^{m}X^{n}$ are unitaries that satisfy $T_{(m,n)}^\dagger=T_{(-m,-n)}$ (again, the factor $2^{-1}$ in the exponent should be understood in the finite field $\mathbb{F}_q$). 
It is apparent in this form that the 
Wigner function is real and normalized, i.e. $\sum_{\bf u} W_\rho({\boldu})=1$. 
Indeed, Wigner functions are quasi-probability distributions of quantum states in phase space. The term ``quasi'' refers to that the Wigner function can take negative value in certain region, which is a convenient indicator of ``quantumness''. 

The displacement operators $T_{\boldu}$ serve as the ``Pauli operators'' of qudit system. For tensored Hilbert space of $N$ qudits, the multi-qudit Pauli operators are defined accordingly 
\begin{equation}
    T_{(m_1,n_1,m_2,n_2,\cdots, m_N,n_N)}=T_{(m_1,n_1)}\otimes T_{(m_2,n_2)}\otimes\cdots \otimes T_{(m_N,n_N)}.
\end{equation}
A stabilizer state in this context is the unique common eigenstate of a set of $N$ commuting Pauli operators. It can be generated by applying Clifford unitaries on computational basis state. 
Clifford unitaries map Pauli operators to Pauli operators by conjugation and  therefore preserve the commutation relation
\begin{equation}
    T_{(m,n)}T_{(p,q)}=w^{mq-np}T_{(p,q)}T_{(m,n)}
\end{equation}
where $mq-np=[(m,n),(n,p)]$ is the standard symplectic form on $\mathbb{F}_q^{2N}$, here $m,n,p,q$ are understood as $N$ dimensional vectors in the context of $N$-qudit systems. 
In this language, the Clifford unitaries act as symplectic transformations $F$ on $2N$ dimensional vector ${\bf u}$ over
$\mathbb{F}_q$ as follows   
\begin{equation}
    UT_{\bf u} U^\dagger=T_{F {\bf u}}.\label{eq:preliminaries_clifford_on_pauli}
\end{equation}
up to phases. Therefore, Clifford unitaries ``rotate'' basis on the phase space and do not change the sign of the Wigner function. 
In fact, a pure state is a stabilizer state if and only if the Wigner function is positive~\cite{Gross_2006}. 
For mixed states with positive Wigner function, it is not necessarily a classical mixture of stabilizer pure states. 
However, the positivity of Wigner function is equivalent to the existence of a set of non-contextual value assignments to all Pauli operators~\cite{Delfosse_2017}, i.e. although the density matrix does not  necessarily look simple, we are still able to reproduce all the measurement outcome via a classical probability distribution~\cite{Veitch_2012,Pashayan_2015}. 

Now, based on the negativity of the Wigner function, one can construct the following magic monotones
\begin{enumerate}
    \item Wigner function one norm and sum negativity
    \begin{equation}
    \lVert\rho\rVert_W :=\sum_{\boldu}\big|W_\rho({\boldu})\big| , \qquad \sn(\rho):=\frac{1}{2}(\lVert\rho\rVert_W-1)
    \end{equation}
    They are magic monotones because that the one-norm is (1) invariant under the Clifford unitaries; (2) non-increasing under partial trace 
and computational basis measurement (upon averaging the one-norm over measurement outcomes).  
For pure states, the sum negativity can give a lower bound to the distance to stabilizer states~\cite{Gross_2021}. 
    \item Mana
    \begin{equation}
    \calM(\rho):=\log\lVert\rho\rVert_W ,
    \end{equation}
    which is also a magic monotone because log is a monotonic and concave function. In addition, it satisfies additivity in the sense that $\calM(\rho\otimes\sigma)=\calM(\rho)+\calM(\sigma)$. In the following section, we will mainly use mana as our magic measure to discuss the dynamics under random circuits. 
\end{enumerate}


Wigner negativity and mana have been applied to many-body systems in \cite{White2020ManaIH,White:2020zoz,Goto:2021anl,Sewell:2022lao,Ahmadi:2022bkg,Tarabunga:2023hau}. 
We also note here that many other measures of magic have been proposed, including stabilizer rank~\cite{Bravyi_2016im,Bravyi_2016tr,Bravyi_2019}, stabilizer extent~\cite{Bravyi_2019} and its extension to mixed states~\cite{Seddon_2021}, robustness~\cite{Howard_2017}, stabilizer nullity~\cite{Beverland:2019jej}, thauma~\cite{Wang_2020}, stabilizer Rényi entropies~\cite{Leone:2021rzd}, etc. The relationship between mana and the other measures have not been fully sorted out. The sum negativity and mana give lower bounds to robustness\footnote{The robustness of magic is defined as 
\begin{equation}
    \calR(\rho)=\min_{x}\left\{\sum_i|x_i|;\ \rho=\sum_ix_i\ket{\psi_i}\bra{\psi_i}\right\}
\end{equation}
where $\ket{\psi_i}$ are stabilizer states. It was shown in \cite{Liu:2020yso} that $\sn(\rho)\leq \frac{1}{2}[\calR(\rho)-1]$ which also implies $\calM(\rho)\leq\log\calR(\rho)$.} of magic and its logarithmic version\cite{Liu:2020yso}.

\section{Haar random circuits} \label{sec:RHC}

As a starter, we first consider the dynamics of Wigner negativity and mana under Haar random circuits. 
We show that the Wigner negativity can be mapped to a special boundary condition in the stat-mech model for the Haar random circuits. This construction offers new physical intuitions on the magic, and its relation to entanglement entropy. Using this setup, We find an interesting competition between magic and entanglement. 

\subsection{The setup and replica trick}

\begin{figure}[t]
    \centering
    \includegraphics[width=0.592\textwidth]{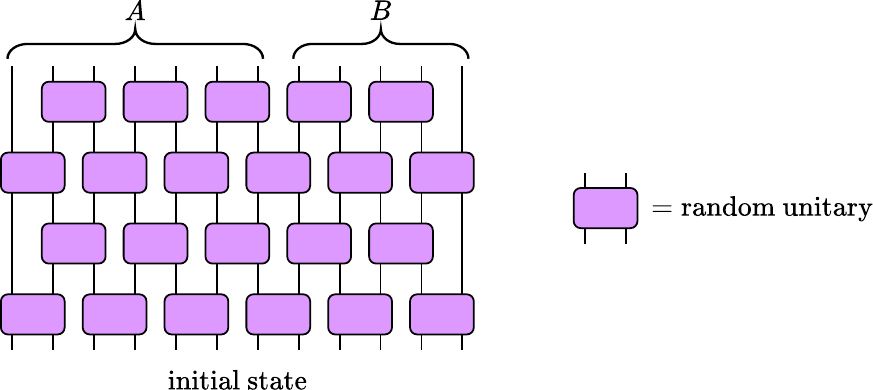}
    \caption{Haar random circuits. We act upon the initial state a circuit that consists of random two-qudit gates arranged in a brickwall pattern. After some depth $t$, we focus on a subregion $A$ and ask how much mana it has. $B$ is the complement region.}
    \label{fig:RHC_setup}
\end{figure}

As shown in Fig.~\ref{fig:RHC_setup}, we are interested in the magic, measured by Wigner negativity and mana, of a subsystem of the output of a Haar random circuit. The initial state $\rho_0$ is left unspecified (later we will see that it is only relevant to the boundary conditions of the stat-mech model). The circuit $U$ consists of Haar random two-qudit gates arranged in a brick-wall pattern. The qudit dimension is an odd prime number $q$. Let $A$ be a subsystem of the output state, whose size is denoted by $|A|$. We investigate how much magic is in $A$, depending on the depth of the circuit $t$.

We employ a replica trick to obtain the Wigner negativity, which involves absolute values that are otherwise hard to handle analytically. The $2n$'th moment of the phase point operators is denoted as
\begin{equation}
    A^{(2n)}:=\sum_{\boldu}A^{\otimes 2n}_{\boldu}.
\end{equation}
We can compute the averaged expectation value of these operators to get the averaged Rényi Wigner negativity, 
\begin{equation}
    W^{(2n)}:=\Ens_U\sum_{\boldu}|W_{\rho_A}(\boldu)|^{2n}=\Ens_Ud_A^{1-2n}\Tr\big[{\textstyle{\frac{1}{d_A}}}A^{(2n)}(U\rho_0U^\dagger)^{\otimes 2n}\big],\label{eq:SM_replica_trick}
\end{equation}
and then perform an analytic continuation to obtain the Wigner function one norm.
\begin{equation}
    \langle\lVert\rho_A\rVert_W\rangle=\lim_{n\rightarrow 1/2}W^{(2n)},
\end{equation}
where $d_A=q^{|A|}$ is the dimension of the Hilbert space associated with  $A$ region, $\Ens_U$ means averaging over the Haar random two-qudit gates.
To get the averaged mana (the ``quenched'' average), an additional replica trick is required:
\begin{equation}
    \langle\calM\rangle=\lim_{k\rightarrow 0}\lim_{n\rightarrow 1/2}\frac{1}{k} W^{(2n,k)}.
\end{equation}
where 
\begin{equation}
    W^{(2n,k)}\equiv\Ens_U \left(d_A^{1-2n}\sum_{\boldu}|W_{\rho_A}(\boldu)|^{2n}\right)^k=\Ens_U\Tr\left[\big({\textstyle{\frac{1}{d_A}}}A^{(2n)}\big)^{\otimes k}(U\rho_0 U^\dagger)^{\otimes 2nk}\right].
\end{equation}
Since the operators that we insert all have the tensor product form, the most favored spin configurations in stat-mech model (will be introduced momentarily) also factorize, i.e. do not connect different $k$-replicas in the $q\rightarrow\infty$ limit. Thus, we simply have
\begin{equation}
    \langle \calM \rangle =   \langle \log \lVert\rho_A\rVert_W  \rangle = \log  \langle\lVert\rho_A\rVert_W\rangle
\end{equation}
for the settings in this paper.

\subsection{Statistical mechanics model}
The ensemble averaging~\eqref{eq:SM_replica_trick} can be done by mapping to a stat-mech model~\cite{Nahum:2017yvy,Zhou:2018myl,Hunter-Jones:2019lps,Bao:2019qah,Jian:2019mny}. Below we give a quick sketch of the procedure. For convenience, we double the replicated Hilbert space and write the Rényi Wigner negativity as
\begin{equation}
    W^{(2n)}=\Ens_U\bra{\tilde{I}_d}({\textstyle{\frac{1}{d_A}}}A^{(2n)}\otimes I)(U^{\otimes 2n}\otimes U^{*\otimes 2n})(\rho_0^{\otimes 2n}\otimes I)\ket{\tilde{I}_d},\quad \ket{\tilde{I}_d}=\sum_{i=1}^{2nd}\ket{i}\ket{i},
\end{equation}
where $d=q^N$ is the total dimension of our system, $\ket{\tilde{I}_d}$ is the (unnormalized) maximally entangled state between the first $2n$ replicas and the other $2n$ replicas.
\begin{figure}
    \centering
    \subfigure[]{\includegraphics[width=0.3\textwidth]{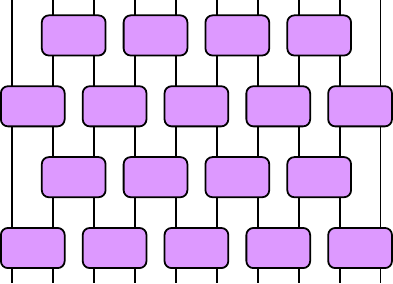}\label{fig:SM_mapping1}} \quad 
    \subfigure[]{\includegraphics[width=0.27\textwidth]{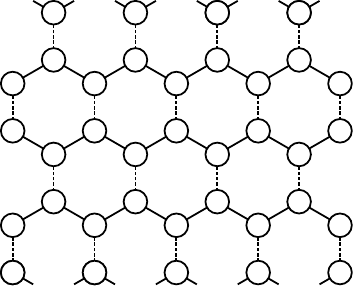}\label{fig:SM_mapping2}} \quad
    \subfigure[]{\includegraphics[width=0.27\textwidth]{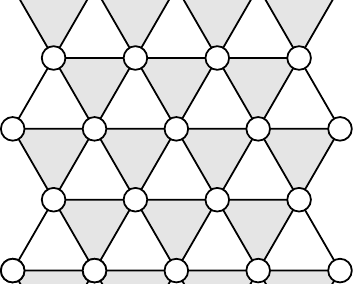}\label{fig:SM_mapping3}}
    \caption{Stat-mech mapping. (a) The random circuit with brickwall structure. (b) Integrating over the Haar random two-qudit gates, we get a honey-comb lattice on which the spins are placed. The spins are permutations in $S_{2n}$. They interact with each other due to the Weingarten functions (the dashed lines) and the contractions (the solid lines). (c) Integrating out the spin at the center of every downward pointing triangle, we get a triangular lattice with three-body weights on downward pointing triangles.}
    \label{fig:SM_mapping}
\end{figure}
Each time we average over a two-qudit unitaries, we get\cite{Collins_2003,Collins_2006}
\begin{equation}
\Ens_U U^{\otimes 2n}\otimes U^{*\otimes 2n}=\sum_{\pi,\sigma\in S_{2n}}\Wg_{q^2}(\pi\sigma^{-1})\big[r(\pi)\otimes I\ket{\tilde{I}_{q}}\bra{\tilde{I}_{q}}r(\sigma^{-1})\otimes I\big]^{\otimes2}.
\label{eq:SM_Haar_weingarten}
\end{equation}
$\pi$ and $\sigma$ are permutations in $S_{2n}$. They are the ``spins'' of the stat-mech model. $\Wg_{q^2}(\pi\sigma^{-1})$ are Weingarten functions in $q^2$ dimensions. Given a $S_{2n}$ permutation $\pi$, the operator associated with it is
\begin{equation}
    r(\pi)=\sum_{\boldx}\ket{\pi\boldx}\bra{\boldx},\quad \ket{\boldx}=\bigotimes_{i=1}^{2n}\ket{x_1}\cdots\ket{x_{2n}},
\end{equation}
where $\ket{x_i}$ is the computational basis of the $i$'th replica. Pictorially, \eqref{eq:SM_Haar_weingarten} says
\begin{equation}
\vcenter{\hbox{\includegraphics[width=0.1\textwidth]{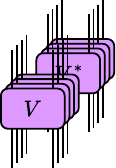}}}=\sum_{\pi,\sigma\in S_{2n}}\Wg_{q^2}(\pi\sigma^{-1})\vcenter{\hbox{\includegraphics[width=0.1\textwidth]{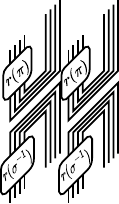}}}.
\end{equation}
This gives a honeycomb lattice shown in Fig \ref{fig:SM_mapping2}, where the dashed lines represent the Weingarten functions. Between one layer and the next, there are contractions
\begin{equation}
\bra{\tilde{I}_{q}}r(\tau^{-1})r(\pi)\otimes I\ket{\tilde{I}_{q}}=\ \vcenter{\hbox{\includegraphics[width=0.05\textwidth]{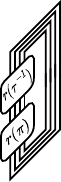}}}\ 
=\Tr[r(\tau^{-1})r(\pi)]=q^{\#(\tau^{-1}\pi)},
\end{equation}
where $\#(\sigma)$ is the number of cycles in the permutation $\sigma$. The coefficients $q^{\#(\tau^{-1}\pi)}$ are represented by solid lines in Fig.~\ref{fig:SM_mapping2}. To proceed, we integrate out the spins in the center of downward pointing triangles to get a triangular lattice shown in Fig.~\ref{fig:SM_mapping3}. In one of these triangles, let $\sigma_1$,$\sigma_2$ be the spins on the top and $\sigma_3$ be the one at the bottom. Integrating out the spin $\tau$ in the center, we get three-body weights
\begin{equation}
J(\sigma_1,\sigma_2;\sigma_3)=\vcenter{\hbox{\includegraphics[width=0.09\textwidth]{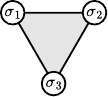}}}=\sum_\tau\vcenter{\hbox{\includegraphics[width=0.09\textwidth]{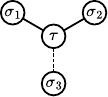}}}=\sum_\tau\Wg_{q^2}(\sigma_3\tau^{-1})q^{\#(\tau\sigma_1^{-1})+\#(\tau\sigma_2^{-1})}.
\end{equation}
At large $q$, the Weingarten functions scale as $\Wg_{q^2}(\sigma_3\tau^{-1})\sim q^{-4n-2|\sigma_3,\tau|}$\cite{Collins_2003,Collins_2006}, where $|\sigma,\tau|$ is a distance function on $S_{2n}$---the minimal number of swaps to turn $\sigma$ into $\tau$. It satisfies $|\sigma,\tau|+\#(\sigma\tau^{-1})=2n$. In other words, the $\tau=\sigma_3$ term gives the leading contribution:
\begin{equation}
    \Wg_{q^2}(\sigma_3\pi^{-1})=q^{-4n}\delta_{\sigma_3\pi}+\text{subleading}.
\end{equation}
In the three body weights, When the spins on the top are aligned ($\sigma_1=\sigma_2$), we have the following:
\begin{equation}
    J(\sigma,\sigma;\pi)=\sum_\tau\Wg_{q^2}(\pi\tau^{-1})q^{2\#(\tau\sigma^{-1})}=\delta_{\sigma\pi}.
\end{equation}
It says that there is no energy cost if the three spins are aligned and that there could not be a horizontal domain wall crossing this triangle. When there is a vertical domain wall ($\sigma_1=\sigma_3\neq\sigma_2$), the weight is
\begin{equation}
    J(\sigma,\pi;\sigma)=\sum_\tau\Wg_{q^2}(\sigma\tau^{-1})q^{\#(\tau\sigma^{-1})+\#(\tau\pi^{-1})}=q^{-|\sigma,\pi|}+\text{subleading}.
\end{equation}
Again the $\tau=\sigma$ term gives the leading contribution. This leads to a $\log q\cdot |\sigma,\pi|$ energy cost.

By far we have determined the bulk of the stat-mech model. On the boundary at the output of the circuit, we have the operator ${\textstyle{\frac{1}{d_A}}}A^{(2n)}$ in \eqref{eq:SM_replica_trick} acting on region $A$. 
This operator is a tensor product of single qudit operators $\frac{1}{q}A^{(2n)}$. (In this subsection we abuse the notation to let $A^{(2n)}$ be on a single qudit). The permutations at the final layer contracts with this operator, so we need to determine which of these permutations are most favored in the large $q$ limit. Any permutation $\pi$ can be broken into cycles with lengths $c_i$ satisfying $\sum_ic_i=2n$.
\begin{equation}
\begin{aligned}
    \Tr\big[{\textstyle{\frac{1}{q}}}A^{(2n)}r(\pi)\big]&=\sum_u\prod_i\Tr(A_u^{c_i})
\end{aligned}
\end{equation}
Using \eqref{eq:phase_point_operators}, we get
\begin{equation}
    \Tr(A_{\boldu}^{c_i})=\Tr(A_0^{c_i})=\begin{cases}
        q,& c_i\text{ even} \\
        1,& c_i\text{ odd}
    \end{cases}
\end{equation}
Therefore, the leading contributions are those permutation made of $n$ swaps. Let $X$ denote such a permutation, then
\begin{equation}
\Tr\big[{\textstyle{\frac{1}{q}}}A^{(2n)}r(X)\big]=q^{-1}\sum_u q^n=q^{n+1}.
\label{eqn: X}
\end{equation}
The subleading terms are suppressed by $\frac{1}{q}$.
The boundary condition on region $B$ is a projection $\ket{\tilde{I}_{q}}$ on every qudit, which essentially sets the boundary spin to be $I$.  
In comparison with $\Tr\big[r(I)\big]=q^{2n}$, \eqref{eqn: X}  is $q^{1-n}$ suppressed. 

This boundary condition for the Wigner negativity is a new ingredient in the stat-mech mapping methodology. 
Later in the next section, for the random Clifford circuits, we will see that ${\textstyle{\frac{1}{q}}}A^{(2n)}$ can be understood as an element of the stochastic orthogonal group (the permutation group is its subgroup). It is a particular spin value in the Clifford stat-mech model. But for now it suffices to know that $X$ is the most favored permutation around it.

To summarize, we obtain the following mapping between the Rényi Wigner negativity and the Replication partition function
\begin{equation}
    \wideboxed{
W^{(2n)}_{\rm Haar} = Z^{(2n)}_{\rm Haar} [A] = \sum_{\sigma_i\in S_{2n}}\prod_{\langle i,j,k\rangle\in \vcenter{\hbox{\includegraphics[width=0.3cm]{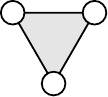}}}}J(\sigma_i,\sigma_j;\sigma_k)
    }
\end{equation}
from which we can obtain the desired magic measure, such as mana of the system. Also see Table~\ref{tab:1} for a summarize of the mapping

\begin{table*}[t]
  \centering
  \begin{tabular}{ |l | l|}
    \hline
   {\bf Haar random circuit} & {\bf Stat-mech model} \\ 
  \hline
 Unitary gates & Permutations (spins)   \\ 
 \hline
  Brick-wall structure & Triangular lattice \\  
  \hline
  Partial trace & Boundary condition with identity (${I}$)\\
    \hline
  Moments of phase point operators & Boundary condition with multi-SWAPs (X)\\ 
    \hline
   Computational basis measurement & Free boundary condition\\ 
    \hline
    Rényi Wigner negativity $W^{(2n)}$&  Replicated partition function\\
  \hline
  Mana & (negative) Free energy \\
    \hline
\end{tabular}
  \caption{A summary of the mapping from the brick-wall Haar random circuit to stat-mech model. 
  }
  \label{tab:1} 
\end{table*}

\subsection{Competition between magic and entanglement}

In the large $q$ limit, the leading contribution to the Rényi Wigner negativity is given by a simple configuration with a domain wall separating $A$ and $B$ ($A$ and its complement $B$ map to different boundary conditions, which force the bulk spins to align to reduce energy).
\begin{figure}[h]
    \centering
    \includegraphics[width=0.44\textwidth]{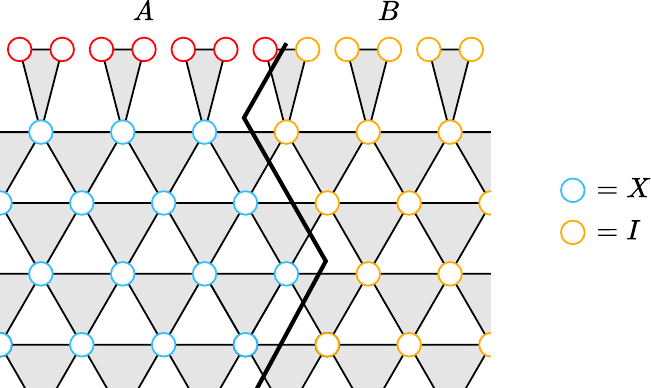}
\end{figure}
Therefore, the Rényi Wigner negativity takes the following simple form 
\begin{equation}
    \log W^{(2n)}=-\log q[(n-1)|A|+n\min l_{A|B}]
\end{equation}
where $l_{A|B}$ is the length of the domain wall separating the domain of spins emanating from $A$ and the domain emanating from its complement $B$. 

A few remarks are in order:
\begin{itemize}
\item The $\log q\cdot (n-1)|A|$ term comes from the the contraction with ${\textstyle{\frac{1}{q}}}A^{(2n)}$ in region $A$. The most favored permutation around it is $X$, but it still comes with a $\log q\cdot(n-1)$ energy cost per site.

\item The $\log q\cdot n\min l_{A|B}$ term comes from the domain wall between $X$ and $I$. Every time the domain wall crosses a downward pointing triangle, it gives an energy cost $\log q\cdot |X,I|=\log q\cdot n$. At large $q$ we should search for the domain wall configuration with smallest length. This procedure is the same one performed in computing the Rényi entropy of $A$, where the boundary condition is replaced by the cyclic permutation $C$ in $A$. The energy cost is given by $\log q\cdot |C,I|=\log q\cdot (n-1)$.

\item The domain wall configurations with minimal energy can be degenerate, but the contribution from the degeneracies is negligible in the $q\rightarrow\infty$ limit. The domain wall between $X$ and $I$ can split into multiple domain walls with the same length, as long as the domains are along a geodesic between $I$ and $X$ in $S_{2n}$. For example it can split into domain walls between $X$ and $\tau$, $\tau$ and $X$ as long as $|X,\tau|+|\tau,I|=|X,I|=n$. These effects are important for finite $q$, but are negligible in $q\rightarrow\infty$ limit\cite{Zhou:2018myl}.
\end{itemize}

Analytically continuing to $n\rightarrow\frac{1}{2}$, we find 
\begin{equation}
    \log\langle\lVert\rho\rVert_W\rangle=\frac{\log q}{2}(|A|-\min l_{A|B}).
\end{equation}

The presence of the $|A|$ term, i.e. volumn law, is because of the final layer of random two-qudit gates. Although these gates are local, they create a lot of magic. The second term is related to entanglement entropy. At large $q$ the entanglement entropy is $S(A)= \log q\cdot \min l_{A|B}$. So we can rewrite the formula as\footnote{Interestingly, this formula also saturates the mana upper bound using the second Rényi entropy (as the Rényi entropy agrees with von Neumann entropy in random circuits at large $q$) \cite{White2020ManaIH}.}
\begin{equation}
\wideboxed{
    \langle\calM\rangle =\frac{1}{2}[\log d_A-S(A)]
    }
    \label{eq: mana haar}
\end{equation}

In conclusion, the final layer of gates is responsible for the volume law ($\frac{\log q}{2}|A|$) contribution to mana. However, the second term implies that the entanglement decreases the magic! 

To better understand this interesting competition between entanglement entropy and magic, let us consider the circuits start with product state (corresponding to free boundary conditions on the initial slice), then the stat-mech model provides  following possibilities for the domain wall cut:
\begin{equation}
\vcenter{\hbox{\includegraphics[width=0.25\textwidth]{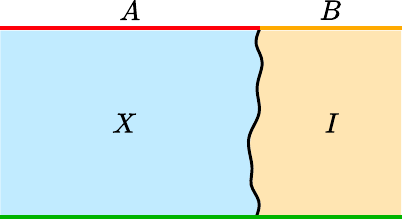}}},\quad\text{or}\quad
\vcenter{\hbox{\includegraphics[width=0.25\textwidth]{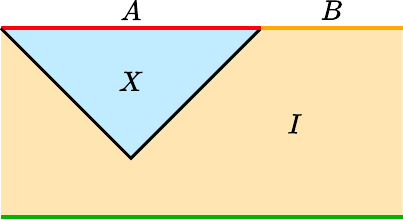}}},\quad\text{or}\quad
\vcenter{\hbox{\includegraphics[width=0.25\textwidth]{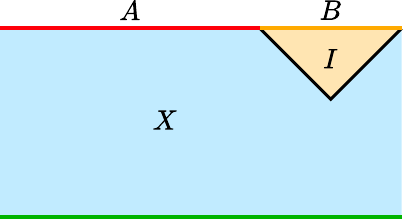}}},
\end{equation}
These configurations have entanglement
\begin{equation}
    \log q\cdot t,\quad \log q\cdot|A|,\quad\text{and}\quad\log q\cdot|B|
\end{equation}
respectively. 
\begin{enumerate}
    \item In the early time,  $t<|A|$, $t<|B|$ case, $S(A)$ grows linearly with $t$, gradually eating up magic.
    \item Late time, and small $A$ case, i.e. $t>|B|>|A|$, the entanglement of $A$ saturates and $A$ is approximately the maximally mixed state with no magic.
    \item Late time, and large $A$ case, i.e. $t>|A|>|B|$, roughly speaking there are $|B|$ many degrees of freedom in $A$ that are maximally mixed, and the remaining $|A|-|B|$ degrees of freedom have mana $\langle\calM\rangle=\frac{\log q}{2}(|A|-|B|)$. 
\end{enumerate}
A heuristic picture for the competition between entanglement and magic can be depicted as follows: for the subregion $A$, there are $S(A)$ degrees of freedom that are maximally mixed with with environment $B$, hence no magic, while the rest of the degrees of freedom are highly magical as the circuit is generated by Haar unitary gates.

\section{Random Clifford circuits}

\label{sec: RCC}

In the last section, we have discussed the scenario where the magic is mainly introduced by the Haar random circuits. In this section, we consider an opposite scenario where the Haar random unitary circuit is replaced by the random Clifford circuit so that the circuit itself does not introduce magic to the final state.  
Instead, the magic will be injected from the initial state or measurements. By this design, we can focus on the spreading and scrambling dynamics of quantum magic. 

We will start with the following setup 
\begin{equation}
\includegraphics[width=0.668\textwidth]{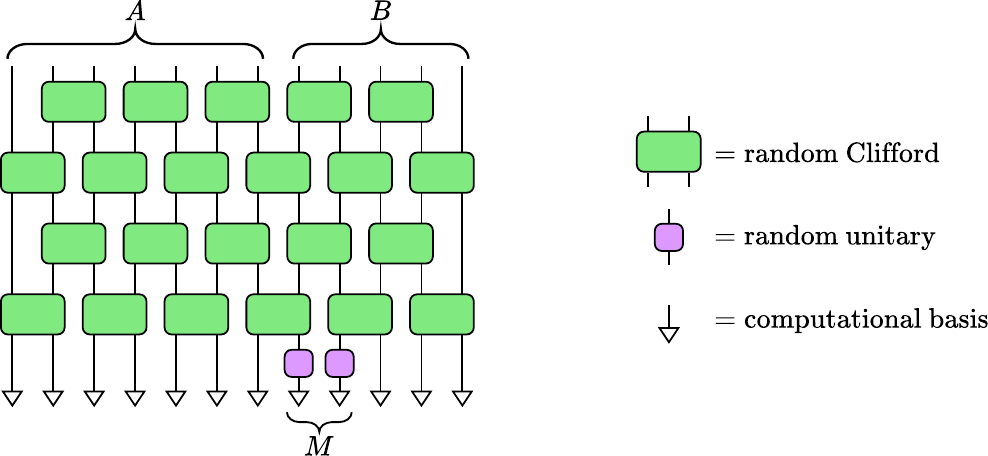} 
\end{equation}
where the main circuit consists of random Clifford two-qudit gates. The magic is introduced by applying single qudit Haar random unitaries at the initial state which is a product of computational basis state. We are interested in the magic measures, i.e. Wigner negativity and mana, in the subregion A of the final state. In this setup, we aim to reveal how the magic spreads and scrambles in many-body states under the random Clifford circuits. 

Next, we consider a setup with computational basis measurements applied to the final state
\begin{equation}
    \includegraphics[width=0.844\textwidth]{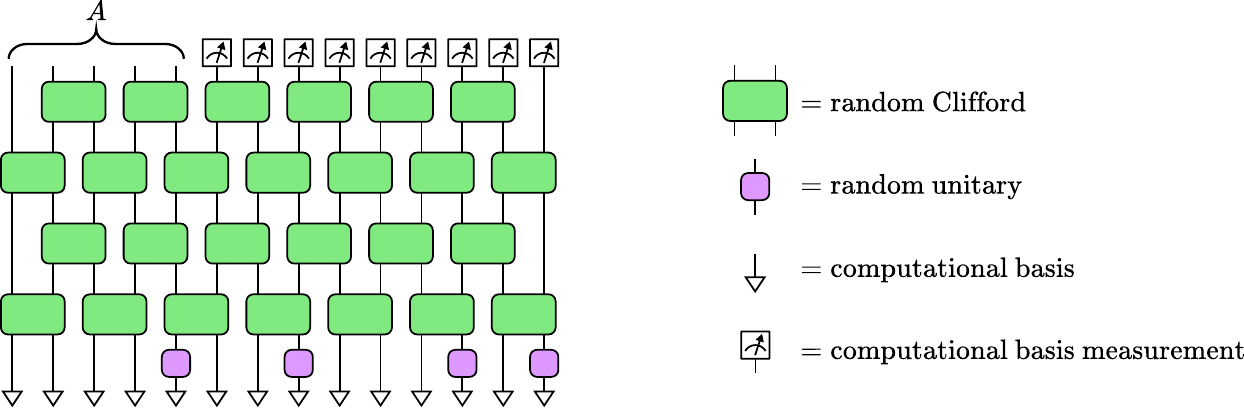} \label{eq:RCC_setup_concentration}
\end{equation}
The goal is to see how the computational basis measurement squeeze the initially sparsed magic.

Now if we instead apply the magical measurement, implemented by single qudit random unitaries before computational basis measurement (equivalent to measuring each qudit in a random basis),
\begin{equation}
    \includegraphics[width=0.736\textwidth]{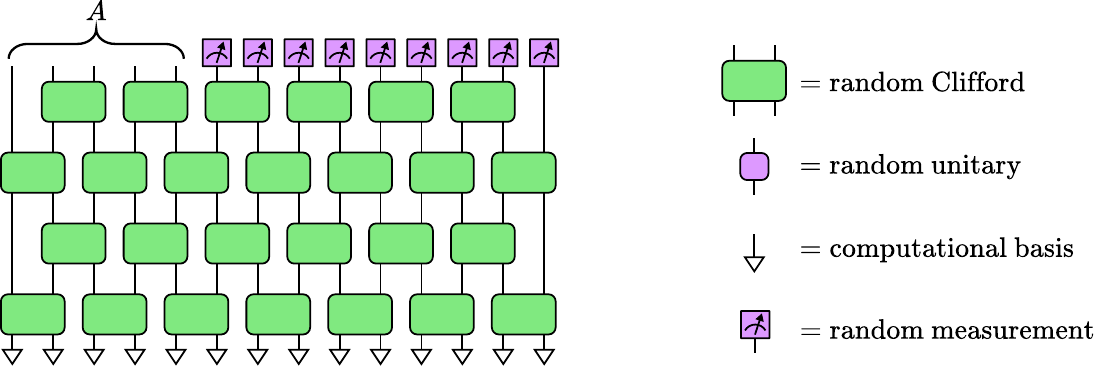}
\end{equation}
we would expect to receive magic teleported from the measurement region. This scenario is, in some sense, similar to the cluster-state measurement based quantum computation (MBQC)~\cite{Raussendorf:2001zim} where the cluster state to start with does not have magic, while the magic can be later injected by non-stabilizer measurement. Note that there is a version of MBQC where magic is in the resource state, and Pauli measurements can achieve universal computation~\cite{Miller_2016}. This scenario is similar to the one depicted in~\eqref{eq:RCC_setup_concentration}.



\subsection{Statistical mechanics model} \label{sec:RCC_SM}

In the stat-mech mapping, the effective ``spin'' degrees of freedom come from the operators in replicated space that commute with the ensemble average.
The average over random Clifford unitaries produces a larger set of spins for the stat-mech model than the one (i.e. permutations) produced under Haar random unitaries in the last section. 
The larger set is known to be the {\it stochastic Lagrangian subspace} \cite{Gross_2021}, whose basis can be denoted as follows
\begin{equation}
    r(T)=\sum_{(\boldx,\boldy)\in T}\ket{\boldx}\bra{\boldy}, \label{clifford_average}
\end{equation}
where $T$ is a stochastic Lagrangian subspace in $\mathbb{F}_q^{4n}$, see appendix~\ref{appendix: clifford average} for details. 

With the extended set of spins, the mapping to the stat-mech follows the case for the Haar random circuits in the previous section with the coupling constants determined by generalized Weingarten functions~\cite{Li:2021dbh}. For the calculation of magic measures related to the Wigner function, we now need to figure out the new boundary conditions
\begin{enumerate}
\item {Boundary condition in the output region}. 
    
The moments of the phase point operators can be written as follows 
\begin{equation}
\begin{aligned}
    {\textstyle{\frac{1}{q}}}A^{(2n)}&={\textstyle{\frac{1}{q}}}\sum_{k,l,\boldx}w^{k\boldx\cdot\boldone}\bigotimes_{i=1}^{2n}\big|l+{\textstyle{\frac{x_i}{2}}}\big\rangle\big\langle l-{\textstyle{\frac{x_i}{2}}}\big|=\sum_{l,\boldx}\delta(\boldx\cdot \boldone)\big|l\bold1+{\textstyle{\frac{\boldx}{2}}}\big\rangle\big\langle l\bold1-{\textstyle{\frac{\boldx}{2}}}\big|
\end{aligned}
\end{equation}
where 
$l\bold1+\frac{\boldx}{2}$ with $\boldx\cdot\bold1=0$ are linear combinations of $\boldone$ and vectors in $\boldone^\perp$. Hence we are summing over all elements of $\mathbb{F}_q^{2n}$. This operator preserves $\ket{\boldone}$ and act as inversions $\ket{\boldx}\rightarrow\ket{-\boldx}$ on those $\boldx\perp\boldone$. It is in fact an stochastic rotation in $\mathbb{F}_q^{2n}$, known as the anti-identity $\bar{I}$.
\begin{equation}
{\textstyle{\frac{1}{q}}}A^{(2n)}=r(\bar{I}),\qquad \bar{I}=n^{-1}\boldone^T\boldone-I
\end{equation}
We have thus found that the boundary condition at $A$ is the anti-identity. Note that in the 2nd Rényi (i.e. $n=1$) case, $\bar{I}=\text{SWAP}$ which is a permutation. 

In Haar circuits, the bulk spins are restricted to permutations, which are quite distant from $\bar{I}$. That is why we have the $q^{1-n}$ suppression when the nearby spin is $X$, giving the extensive contribution $\frac{\log q}{2}|A|$ to mana in the replica limit. In the Clifford case, we get more spin elements, including $\bar{I}$. When the nearby spin is $\bar{I}$, there are no energy cost, hence no extensive contribution to mana.  \emph{This gives a stat-mech perspective on the fact that pure Clifford circuits do not increase magic.} 

\item { Boundary condition at the magic insertions by Haar random unitaries.}

If the magic is injected by single qudit Haar random unitaries, it will average to permutations. Then the contraction $\Tr[r(\pi) r(\bar{I})]$ gives rise to the interactions between $\bar{I}$ and the permutations. It is given by the number of solutions to the equation $\pi\bar{I}\boldx=\boldx$. Equivalently
\begin{equation}
    (\pi+I)\boldx=n^{-1}(\boldx\cdot\boldone).
\end{equation}
Using the decomposition $\boldx=l\boldone+\boldy$ where $\boldy\perp\boldone$, the condition becomes $\pi\boldy=-\boldy$. It has solutions when $\pi$ consists of cycles with lengths no greater than $2$. If $\pi$ has is such a permutation with $m$ swaps, then there are $q^{m}$ possible solutions of $y$. Taking into account the $q$ possibilities of $l$, we get
\begin{equation}
\Tr[r(\pi) r(\bar{I})]=q^{m+1}.
\end{equation}
The $m=n$ case is the leading piece with $\Tr[r(X) r(\bar{I})]=q^{n+1}$. 

Notice that if the magic is injected by single-qudit random unitaries, then the spins in $M$ can differ form site to site. If we do a multi-qudit random unitary instead, the spins are forced to be the same permutation at different sites.

\item {Boundary condition at computational basis}

For computational basis state $\ket{x}\bra{x}^{\otimes 2n}=\ket{x\bold1} \bra{x\bold1} $ and any spin $T$ i.e. element of stochastic Lagrangian subspace, we have 
\begin{equation}
    \Tr\big[r(T)\ket{x}\bra{x}^{\otimes 2n}\big]=\bra{x\bold1}r(T)\ket{x\bold1}=1.\label{eq:SM_free_boundary_condition}
\end{equation}
where we used $r(T)\ket{x\bold1}=\ket{x\bold1}$ because $T$ is stochastic. Since all stochastic rotations give the same number, the boundary condition is free for both initial state and final state measurement. 
\end{enumerate}



\begin{table*}[t]
  \centering
  \begin{tabular}{| l | l|}
    \hline
  {\bf  Random Clifford circuit }& {\bf Stat-mech model} \\ 
  \hline
 Unitary gates & Stochastic Lagrangian subspace (spins)   \\ 
 \hline
  Brick-wall structure & Triangular lattice \\  
  \hline
  Partial trace & Boundary condition with identity (${I}$)\\
    \hline
  Moments of phase point operators & Boundary condition with anti-identity ($\bar{I}$)\\
  \hline
  Haar random unitaries (magic injection) & Boundary condition permutations ($X$ or $I$)\\
  \hline
   Computational basis measurement & Free boundary condition\\ 
    \hline
    Rényi Wigner negativity $W^{(2n)}$&  Replicated partition function\\
  \hline
  Mana & (negative) Free energy \\
  \hline
\end{tabular}
  \caption{A summary of the mapping from the brick-wall random Clifford circuit to stat-mech model. 
  }
  \label{tab:2} 
\end{table*}

In parallel to the Haar random case, we can summarize the mapping as follows (also see Table~\ref{tab:2}) 
\begin{equation}
    \wideboxed{
W^{(2n)}_{\rm Clifford} = Z^{(2n)}_{\rm Clifford} [A] = \sum_{T_a\in\Sigma_{2n}(q)}\prod_{\langle a,b,c\rangle\in \vcenter{\hbox{\includegraphics[width=0.3cm]{Figure/SM_downward_pointing_triangle.pdf}}}}\tilde{J}(T_a,T_b;T_c)
    }
\end{equation}
where the summation runs over $\Sigma_{2n}(q)$ which is the set of stochastic Lagrangian subspaces $T\subseteq \mathbb{F}^{4n}_q$. $\tilde{J}(T_a,T_b;T_c)$ is the three body weight obtained from the Clifford average\cite{Li:2021dbh} (see Appendix \ref{appendix: clifford average} for a review of the derivation), with the following properties similar to those of the Haar counterpart:
\begin{equation}
    \tilde{J}(T_a,T_a;T_b)=\delta_{T_a,T_b},\qquad \tilde{J}(T_a,T_b;T_a)=q^{-|T_a,T_b|}+\text{subleading}.
\end{equation}

To obtain the Rényi Wigner negativity and related mana from the stat-mech model, we need to find the leading contributions of all possible spin configurations.  
At large $q$, the leading contributions are given by the spins $I$, $\bar{I}$ and $X$. The relevant distances are
\begin{equation}
\label{eq:distance}
    |\bar{I},I|=2n-1.\quad |\bar{I},X|=n-1,\quad |X,I|=n.
\end{equation}
Let $l_{\bar{I},X}$, $l_{\bar{I},I}$, and $l_{X,I}$ denote the lengths of the domain walls between $\bar{I}$ and $X$, $\bar{I}$ and $I$, and $X$ and $I$, respectively. Then the Rényi Wigner negativity is given by
\begin{equation}
\log W^{(2n)}=-\log q\cdot \min\big[(n-1)l_{\bar{I},X}+(2n-1)l_{\bar{I},I}+nl_{X,I}\big]. \label{eq:RCC_renyi_negativity_mincut}
\end{equation}

\subsection{Magic dynamics: spreading and scrambling}

Now, we consider the first setup that at $t=0$ we inject magic by applying Haar random unitaries in region $M$. Then the random Clifford circuits spreads and scrambles the magic.
\begin{equation}
    \includegraphics[width=0.32\textwidth]{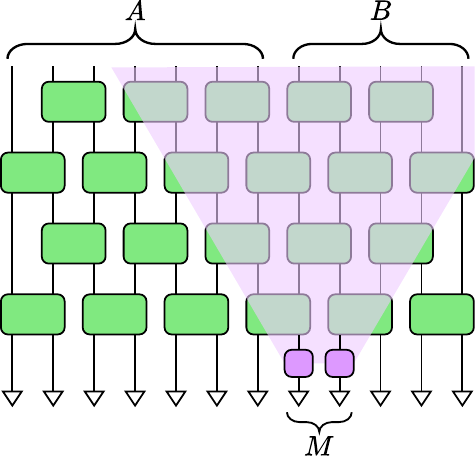}
\end{equation}
We divide the whole system into $A$, $B$ and ask how much magic is in $A$. We obtain the following results using the stat-mech model 
\begin{enumerate}
    \item Early time region. Here we assume $|A|>|B|$ for simplicity, and early time refers to  $t<|A|$.

At early time, 
Let $J^+(M)$ denote the causal future of $M$---the union of the lightcones of points in $M$ determined by the brickwall circuit structure. If $A\supseteq J^+(M)$, then the magic should be all inside $A$. 
Indeed, from the stat-mech perspective, the $\bar{I}$ domain connecting $A$ is in contact with the $X$ domain connecting $M$ due to the causual structure, therefore 
\begin{equation}
\log W^{(2n)}=\min\left\{\vcenter{\hbox{\includegraphics[width=0.215\textwidth]{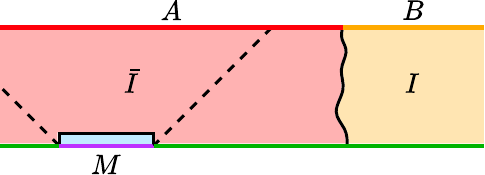}}},\quad \vcenter{\hbox{\includegraphics[width=0.215\textwidth]{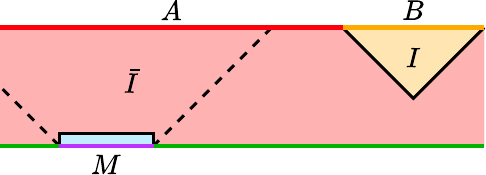}}}\right\}.
\end{equation}
In the replica limit, both configurations give
\begin{equation}
    \langle\calM\rangle=\frac{\log q}{2}|M|,
\end{equation}
meaning that $A$ takes all the initial magic independent of $t$. 

On the other hand, if the causal future $J^+(M) \subseteq B $, i.e. the magic does not reach $A$. The state in $A$ can thus be purified by a stabilizer state with no magic, consistent with the stat-mech result that $\langle\calM\rangle =0$ as the $\bar{I}$ domain cannot reach $M$ due to the lightcone structure.

A somewhat special scenario is when 
$J^+(M)$ is neither covered by $A$ nor $B$. Let $J^-(A)$ be the causal past of $A$. If $t$ is small such that $J^-(A)$ intesects $M$, then the $\bar{I}$ domain cannot just avoid $M$. In $J^-(A)\cap M$, the most favorable permutation is $X$. For the region in $M$ that is outside $J^-(A)$, the spin can simply be $I$ and has no energy cost when it meets the $I$ domain. 
\begin{equation}
\begin{aligned}
\log W^{(2n)}&=\min\left\{\vcenter{\hbox{\includegraphics[width=0.215\textwidth]{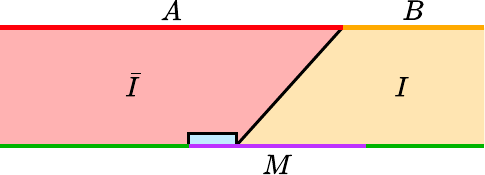}}},\quad \vcenter{\hbox{\includegraphics[width=0.215\textwidth]{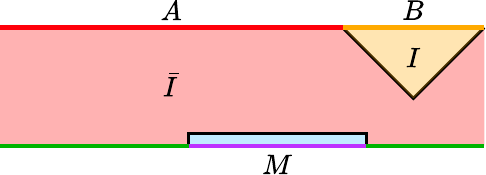}}}\right\} \\
&=\log q\cdot\min\{n|J^-(A)\cap M|+(2n-1)t,\ n|M|+(2n-1)|B|\}.
\end{aligned}
\end{equation}

To better understand the above formula, let us consider the following simplifying scenario
\begin{enumerate}
    \item If we further assume $t<|B|$, we have
    \begin{equation}
\log W^{(2n)}=\log q\cdot\big[n|J^-(A)\cap M|+(2n-1)t\big],\qquad \langle\calM\rangle=\frac{\log q}{2}|J^-(A)\cap M|.
\end{equation}
That is to say, the magic comes from the part of $M$ that is covered by the backward of lightcone of $A$.

\item While for $t>|B|$ (but still $t<|A|$), i.e. $B$ is completely mixed, if we take $|M|\ll t-|B|$ then
\begin{equation}
\log W^{(2n)}=\log q\cdot\big[n|M|+(2n-1)|B|\big],\qquad \langle\calM\rangle=\frac{\log q}{2}|M|.
\end{equation}
That is to say, 
$A$ will take all the magic even though its backward lightcone does not cover $M$. 
\end{enumerate}

\item 
Late time region, i.e. $t\geq |A|$ and $t\geq |B|$.  
    
At late time, 
we can ignore the vertical minimal cuts, then we have 
\begin{equation}
\log W^{(2n)}=\min\left\{\vcenter{\hbox{\includegraphics[width=0.18\textwidth]{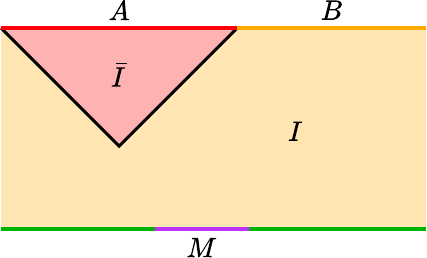}}},\quad
\vcenter{\hbox{\includegraphics[width=0.18\textwidth]{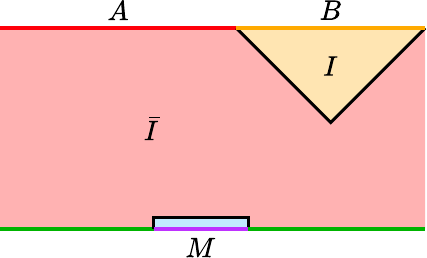}}},\quad
\vcenter{\hbox{\includegraphics[width=0.18\textwidth]{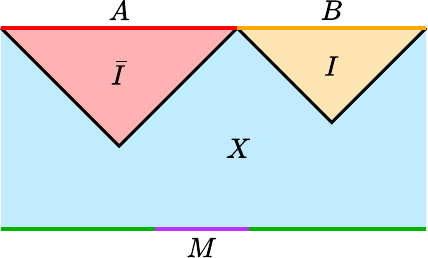}}}\right\}.
\end{equation}
\begin{equation}
\langle\calM\rangle=\frac{\log q}{2}\cdot\begin{cases}
    0,& |A|-|B|<0 \\
    |M|,& |M|<|A|-|B| \\
    |A|-|B|,& |M|>|A|-|B|>0
\end{cases}\label{eq:initial_magic_infinite_depth}
\end{equation}
Since random Clifford unitaries form a 2-design\cite{Zhu_2017,Webb:2015vwz} (for qubits, they are 3-design), Page's theorem indicates that the smaller subregion will be maximally mixed. That is an intuition why we do not obtain any magic if $|A|<|B|$.  
Now for $|A|>|B|$, a picture to understand the result it that there are $|B|$ many degrees of freedom in $A$ that form EPR pairs with $B$. Those degrees of freedom remain maximally mixed, giving no contribution to magic. The remaining $|A|-|B|$ degrees of freedom contribute to ``host'' the magic. If $|M|<|A|-|B|$ then these degrees of freedom captures all the initial magic. If $|M|>|A|-|B|$ then they can only capture as much magic as possible, which is $|A|-|B|$.
\end{enumerate}

To summarize, the picture here is that for shallow circuit, i.e. the circuit depth is smaller than the size of subregions in question, the mana is determined by the ``causal diamond'' of the output state. While for deep circuit, some of the region has been fully scrambled in terms of entanglement entropy, the mana will be transfered to the remaining regions, upbounded by the number of degrees of freedom that is still un-entangled with the environment, i.e. $|A|-|B|$ in our setting when $|A|$ is larger.



\subsection{Measurement effects: concentration and teleportation}


Another fundamental operation in quantum circuits is measurement. It is the ``output'' of a quantum processor. Measurement also plays a pivot role in the quantum error correction. In this subsection, we investigate the measurement effects on quantum magic. We consider two types: (1) computational basis measurement and (2) ``magical'' measurement, more specifically measurement in Haar random basis 
\begin{enumerate}
    \item {\bf Magic concentration.} For the first one, we inject magic in the initial state and perform measurement on certain region of the outcome state and measure the magic of the remaining region shown as follows
    \begin{equation}
    \begin{aligned}
            \includegraphics[width=0.376\textwidth]{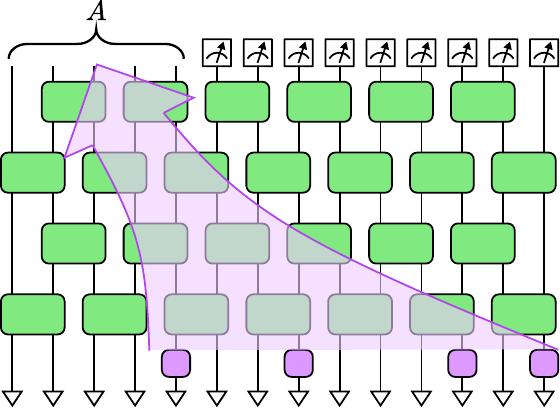}
    \end{aligned}
    \end{equation}
    More explicitly, here we add random unitaries on some sites at the first layer of the circuit that could be sparsely distributed. Then at time $t$, we measure all the complement of $A$ in the computational basis. We compute the mana in $A$ averaged over the measurement outcomes and the random gates. We can do the gate average first for a fixed measurement outcome, and then average over measurement outcomes afterwards. For fixed outcome, we get free boundary conditions in the measured region according to \eqref{eq:SM_free_boundary_condition}. Then we find that the result is independent of the outcome, so the outcome average can be done trivially.
\begin{equation}
\begin{aligned}
\log W^{(2n)}&=\min\left\{\vcenter{\hbox{\includegraphics[width=0.2\textwidth]{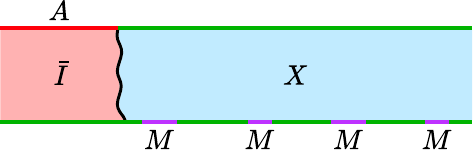}}},\ \vcenter{\hbox{\includegraphics[width=0.2\textwidth]{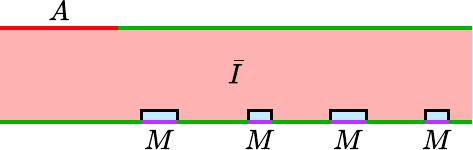}}},\ 
\vcenter{\hbox{\includegraphics[width=0.2\textwidth]{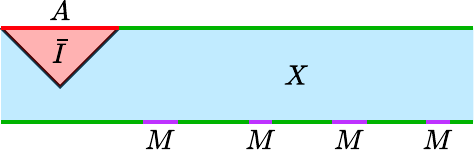}}}\right\} \\
&=-(n-1)\log q\cdot\min\{t,\ |M|,\ |A|\}, \\
\langle\calM\rangle&=\frac{\log q}{2}\cdot\min\{t,\ |M|,\ |A|\}.
\end{aligned}
\end{equation}
At small $t$, the mana is bounded by $t$. For $|M|<|A|$, all mana is concentrated to $A$ at $t>|M|$. If $|M|>|A|$, then $A$ cannot take in all the mana---the mana saturates after $t=|A|$.

The lesson is that magic tends not to be destroyed by computational basis measurements, but to be squeezed into the unmeasured region. Starting from very ``sparse'' magic, computation basis measurements on qubits outside of $A$ can concentrate the magic to $A$. However, the concentration capacity is bounded by the circuit depth (i.e. $t$ here). In order to fully extract the magic injected in the initial state, we require a random Clifford circuit of $O(|A|)$ depth that scrambles the system.

 \item {\bf Magic teleportation.} In the second case, we start with the initial product state of computational basis with no magic and evolve under Clifford unitaries. But in the final state, we perform magical measurement on certain region and measure the consequence in the remaining region. 
\begin{equation}
\begin{aligned}
        \includegraphics[width=0.376\textwidth]{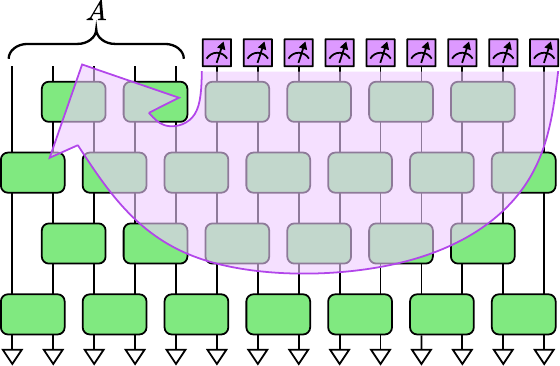}
\end{aligned}
\end{equation}
More explicitly, the magical measurement is modeled by a Haar random unitary followed by a computational basis measurement. The magical measurements is performed in region $M$ that is the complement of $A$.  

\begin{equation}
\begin{aligned}
\log W^{(2n)}&=\min\left\{\vcenter{\hbox{\includegraphics[width=0.17\textwidth]{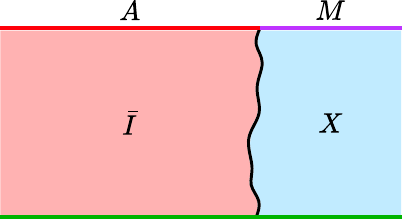}}},\quad \vcenter{\hbox{\includegraphics[width=0.17\textwidth]{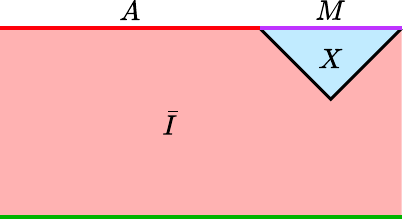}}},\quad 
\vcenter{\hbox{\includegraphics[width=0.17\textwidth]{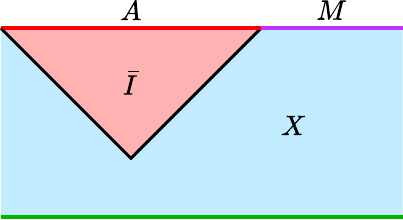}}}\right\} \\
&=-(n-1)\log q\cdot\min\{t,\ |M|,\ |A|\}, \\
\langle\calM\rangle&=\frac{\log q}{2}\cdot\min\{t,\ |M|,\ |A|\}.
\end{aligned}
\end{equation}
The mechanism behind the magic injection is teleportation. Before the measurements, the amount of entanglement between $A$ and its complement provides the resource for teleportation. In fact, the pre-measurement entanglement is exactly $\log q\cdot\min\{t,\ |M|,\ |A|\}=2\langle\calM\rangle$. 

\end{enumerate}

\subsection{Relationship between magic injection and coherent information}

If the magic is injected in the initial state, we can think of the process of sending magic through some stabilizer channel from $M$ to $A$. The channel can be written as
\begin{equation}
    \calE(\rho)=\Tr_B(U\rho\otimes\ket{0}_{\bar{M}}\bra{0}_{\bar{M}}U^\dagger),
\end{equation}
where $\ket{0}_{\bar{M}}$ is the initial product state on the complement of $M$. The ability of the channel to send magic should be bounded by its ability to send quantum information, which is quantified by \emph{coherent information}~\cite{Schumacher:1996dy,Lloyd:1996at}. Let us imagine maximally entangling $M$ with a reference $R$ whose size is same as $M$.
\begin{equation}
\vcenter{\hbox{\includegraphics[width=0.36\textwidth]{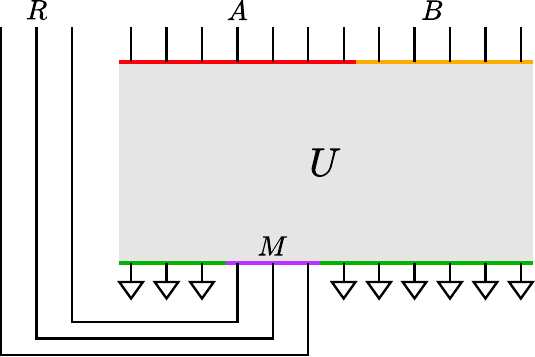}}}
\end{equation}
The coherent information is given as 
\begin{equation}
    I_c(R;A)=S(A)-S(AR).
\end{equation}
It can also be computed using the stat-mech mapping for Rényi entropies~\cite{Gullans:2020whl} combined with a replica trick. In computing the Rényi entropy $S^{(n)}(A)=\frac{1}{1-n}\log\Tr(\rho_A^n)$, we should insert cyclic permutation operator at $A$, which fixes the boundary spins to be the cyclic permutation $C$. In region $B$ that was traced out, the boundary condition is $I$. It is also $I$ in $M$, because in the initial state $M$ is maximally mixed. The entropy comes from the domain wall energy between $C$ and $I$. The domain wall should separate the domains emanating from $A$ and $B\cup M$. Suppose it has length $l_{A|BM}$. We thus have $\langle S^{(n)}(A)\rangle=\log q\cdot \min l_{A|BM}$, where we used $|C,I|=n-1$. In computing the Rényi entropy $S^{(n)}(AR)$, the boundary condition in $M$ becomes $C$. As a consequence, the domain wall should separate $A\cup M$ and $B$. Suppose it has length $l_{AM|B}$, then $\langle S^{(n)}(AR)\rangle=\log q\cdot \min l_{AM|B}$. In conclusion,
\begin{equation}
    \langle I_c(R;A)\rangle=\log q\cdot (\min l_{A|BM}-\min l_{AM|B}), 
\end{equation}
which coincidents with the formula for mana with multi-qudit random unitaries injection! That is to say, in this case (see Appendix \ref{appendix: details of the energy minimization} for more details)
\begin{equation}
\wideboxed{
    \langle\calM\rangle=\frac{1}{2}\max\{\langle I_c(R;A)\rangle,\ 0\}
    }
    \label{eq: coherent}
\end{equation}

In the magic concentration section, we perform computational measurements on the final state and compute the mana averaged over measurement outcomes.
\begin{equation}
\vcenter{\hbox{\includegraphics[width=0.36\textwidth]{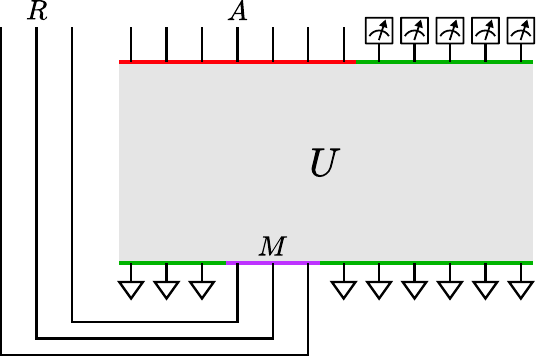}}}
\end{equation}
For a fixed measurement outcome, we think of the process as a map from $M$ to $A$, which might not be a channel. Nonetheless, we can compute $\tilde{I}_c(R;A)\equiv S(A)-S(AR)$. Again we have $\langle\calM\rangle=\frac{1}{2}\max\{\langle \tilde{I}_c(R;A)\rangle,\ 0\}$.

If magic is injected by the magical measurements in the final state, we need to consider the following process
\begin{equation}
\vcenter{\hbox{\includegraphics[width=0.49\textwidth]{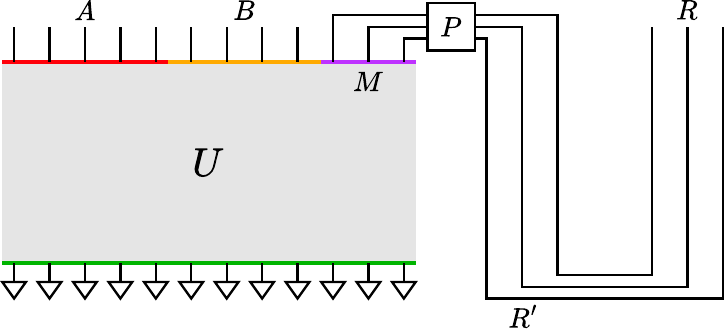}}}
\end{equation}
Initially we maximally entangle the reference $R$ with $R'$. After the Clifford unitary $U$, we measure the multi-qudit Bell basis in $M$ and $R'$. Depending on the measurement outcome, we get an additional Pauli operator $P$. These operators doesn't affect the result, so again we have $\langle\calM\rangle=\frac{1}{2}\max\{\langle \tilde{I}_c(R;A)\rangle,\ 0\}$. 

To summarize, the magic dynamics studied in the setting of this paper can be precisely related to the coherent information of the corresponding channel by a simple formula \eqref{eq: coherent}. 
The stat-mech mapping provides the bridge to connect these two seemingly different concepts. It will be interesting to figure out if this relation (or weaker version) holds in a more general setting, i.e. without assuming large local dimensional of Hilbert space etc.


\section{Summary and discussions}
\label{sec: summary}

In this work, we study the dynamics of magic, characterized by Wigner negativity and mana, under random circuits. 
We derived the map of the Wigner negativity under average (Haar and Clifford) to 
stat-mech model, in particular the 
boundary conditions that are responsible for computing the negativity. 
Using this framework, we obtain exact results regarding the magic structure of a highly entangled many-body state coming from random circuits at large local Hilbert space dimension.

\begin{enumerate}
    \item First of all, we observe an interesting quantitative competition between magic and entanglement (c.f. \eqref{eq: mana haar})
    \begin{equation}
        \langle\calM\rangle =\frac{1}{2}[\log d_A-S(A)]
    \end{equation}
    which is derived for a state prepared by Haar random unitaries. However, the message here should be general: magic of a subregion 
    is limited if such region is highly entangled with the environment. 
    \item For random Clifford circuit, we obtain precise formulas for spreading and scrambling of magic. With measurements, we find interesting phenomena of magic squeezing and teleportation, depicted as follows
    \begin{figure}[h]
    \centering
    \raisebox{-.55\height}{\includegraphics[width=0.213\textwidth]{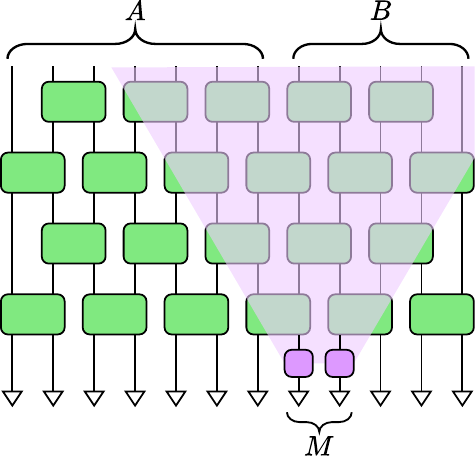}}
    \qquad
    \raisebox{-.5\height}{\includegraphics[width=0.25\textwidth]{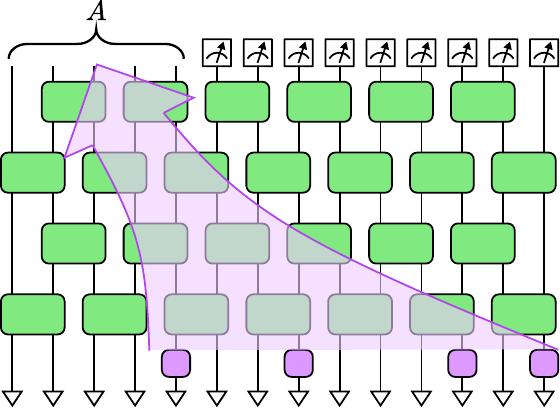}}
    \qquad
    \raisebox{-.45\height}{\includegraphics[width=0.25\textwidth]{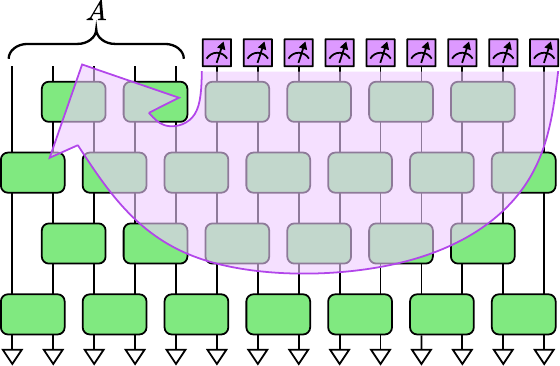}}
    \end{figure}
    
    It is also worth to mention that the capacity of magic squeezing and teleportation is constrained by the random circuit depth, vivid from the above illustration. 
    \item Based on the above magic dynamics, we find an interesting connection to the coherent information 
    (c.f. \eqref{eq: coherent})
    \begin{equation}
    \langle\calM\rangle=\frac{1}{2}\max\{\langle I_c(R;A)\rangle,\ 0\}
    \end{equation}
    here, we regard the circuits as channels to send information from region $M$ (which is maximally entangled with a reference $R$) to $A$. The basic observation is that, in our setting with large local Hilbert space and random circuits, the magic sent through the channel saturates the bound given by the coherent information.  
\end{enumerate}
An enhanced understanding of the ``magic'' dynamics within many-body systems holds potential benefits not only for the quantum computation community but also for the field of condensed matter physics. Beyond uncovering deeper layers of quantum correlations within phases of matter, such understanding could also pave the way for the development of new, resource-efficient computational algorithms, for example the Clifford-augmented tensor algorithms~\cite{Lami:2024osd,Qian:2024vea,Qian:2024wun,Mello:2024jev,Qian:2024lqx}.

For future, there are a lot of fundamental aspects of magic to explore in this framework
\begin{enumerate}
\item In this paper, the relation between magic and entropic quantities, such as the above mentioned two formulas (c.f.\eqref{eq: mana haar} and \eqref{eq: coherent}) are dervied in certain limit. It will be interesting to obatin more general equalities or inequalities relating magic and entanglement~\cite{Zhou_2020,True:2022ypf,Tirrito:2023fnw,Cao:2024nrx,Gu:2024qvn}.

    \item Noise effect in the circuit~\cite{Li:2021par,Li:2022ycx,Dias:2022goz,Li:2023ekc,Liu:2024pyz}? This is close to the original study on the magic state distillation~\cite{PhysRevA.71.022316,knill05};
    \item Relation to Chaos~\cite{Leone_2021,Goto:2021anl,Ahmadi:2022bkg,Passarelli:2024lpm}? Both can be studied in the random unitary circuits, therefore, quantitative connection may be built in this language. 
    \item Field theory analog? Replica trick and twist operators have been successful in the computation of entanglement entropy for quantum field theories. Can magic be defined and computed in field theory? 
    \item What is the role of magic in holography? Can entanglement alone (without magic) build up spacetime? In fact, the techniques developed in this work can be directly applied to study the magic distribution in random Clifford tensor networks which mimic bulk geometry. For example, if one inject magic at a bulk point (i.e. by making one of the tensors Haar random), then a boundary region receives this magic if the bulk point is in the entanglement wedge.
\end{enumerate}

\section*{Acknowledgement}

We thank
Yaodong Li,
Zi-Wen Liu, 
Yifei Wang,
Zhi-Cheng Yang 
for discussions. We thank Xiao-Liang Qi for discussions and suggesting the potential connection with coherent information.
We thank Yimu Bao for comments on the draft.
We acknowledge support from 
the National Key R\&D Program of China 2023YFA1406702, 
the Tsinghua University Dushi Program 
and the DAMO Academy Young Fellow program.

\appendix

\section{Comparison with Stabilizer Rényi entropy}
\label{appendix: SRE}


We can also use the stat-mech mapping to compute quantities based on the Pauli expansion of the state, captured by the Weyl function. The Weyl function is defined as the expectation value of Pauli operators:
\begin{equation}
    \chi_\rho(\boldu)=\Tr(T_{\boldu}\rho).
\end{equation}
The distribution of their absolute values are also known as the Pauli spectrum~\cite{Beverland:2019jej}, which captures the stabilizer Rényi entropy (SRE)\cite{Leone:2021rzd,PhysRevLett.132.240602}, stabilizer norm\cite{Campbell_2011} and magic gap\cite{Bu:2023ssg,Bu:2023uic}. Ref ~\cite{Turkeshi:2023lqu} studied the Pauli spectrum in Haar random pure states and obtained numerical results for pure states generated by long depth random Haar circuits.

To access the Pauli spectrum, we can work out the moments $\sum_{\boldu}|\Tr(T_{\boldu}\rho)|^{2n}$. They can be written as the expectation value of moment operators
\begin{equation}
    T^{(2n)}\equiv\sum_{k,l}T_{(k,l)}^{\otimes n}\otimes T_{{(k,l)}}^{\dagger\otimes n}.
\end{equation}
They play a similar role as $A^{(n)}$ in computing the moments of the Wigner function. For a single qudit,
\begin{equation}
    T_{(k,l)}=w^{-kl/2}Z^kX^l=\sum_xw^{-kl/2}w^{k(x+l)}\ket{x+l}\bra{x}=\sum_x w^{kx}\big|x+{\textstyle{\frac{l}{2}}}\big\rangle\big\langle x-{\textstyle{\frac{l}{2}}}\big|.
\end{equation}
\begin{equation}
\begin{aligned}
    {\textstyle{\frac{1}{q}}}T^{(2n)}&={\textstyle{\frac{1}{q}}}\sum_{k,l}\sum_{\boldx,k,l}w^{k(\boldx\cdot\bolds)}\big|\boldx+{\textstyle{\frac{l\bolds}{2}}}\big\rangle\big\langle \boldx-{\textstyle{\frac{l\bolds}{2}}}\big|=\sum_{l,\boldx}\delta(\boldx\cdot\bolds)\big|\boldx+{\textstyle{\frac{l\bolds}{2}}}\big\rangle\big\langle \boldx-{\textstyle{\frac{l\bolds}{2}}}\big| 
\end{aligned}
\end{equation}
where $\bolds=(\underbrace{1,\cdots,1}_{n},\underbrace{-1,\cdots,-1}_{n})$. This operator corresponds to a stochastic rotation $S\in O_{2n}(q)$:
\begin{equation}
    {\textstyle{\frac{1}{q}}}T^{(2n)}=r(S),\qquad S=I-n^{-1}\bolds\bolds^T.
\end{equation}
At $n=1$, $S$ is swap. The permutation that is nearest to $S$ is the identity $I$, with distance $|S,I|=1$. In the stat-mech model, these operators fix the boundary spins in region $A$ to be $S$. In the complement region $B$, the spins are fixed to be $I$. Since $I$ is permutation with the smallest distance to $S$, $I$ and $S$ will be the relevant spins in the model, while the other spins value are not important at large $q$.

As an application, we compute the stabilizer Rényi entropies. The stabilizer 2-entropy is defined as \cite{Leone:2021rzd}
\begin{equation}
    M_2(\rho)=-\log\frac{\sum_{\boldu}|\chi_\rho(\boldu)|^4}{\sum_{\boldu}|\chi_\rho(\boldu)|^2}=-\log\frac{\sum_{\boldu}|\Tr(T_{\boldu}\rho)|^4}{d_A\Tr(\rho^2)}
\end{equation}
For pure or mixed stabilizer states, $|\chi_\rho(\boldu)|$ is either $0$ or $1$ which means $M_2(\rho)=0$. So it is a witness for magic in this sense. 

In the Clifford stat-mech model, $I$ will be the most favorable permutation in the magic injection region. The domain wall between $S$ and $I$ is the one separating the domains emanating from $A$ and $B\cup M$. In the $q\rightarrow\infty$ limit, we have the simple minimal cut formula:
\begin{equation}
\langle M_{2n}\rangle=\log q\cdot \min l_{A|BM}-S(A). \label{eq:RCC_SRE_mincut}
\end{equation}
This formula implies that stabilizer Rényi entropy cannot distinguish between magic injected by multi-qudit random unitaries and magic injected by single-qudit random unitaries. This follows from the fact that the stat-mech model for SRE, all spin elements should be identity in the magic injection region to lower energy cost. In contrast, in the mana computation, the spins in $M$ can take either $X$ or $I$ if the magic is due to single-qudit random unitaries. In addition, the second term of \eqref{eq:RCC_SRE_mincut} differs from that of \eqref{eq:RCC_multiquditrandom_mana_mincut}. In the mana computation, $l_B$ is the cut that separates $AM$ and $B$, while in the SRE case the cut for $S^{(2)}$ only has to separate $A$ and $B$---it does not care about $M$. So $S^{(2)}(A)\leq \log q\cdot l_{AM|B}$ in the $q\rightarrow\infty$ limit. It follows that SRE is always bigger than mana, when the magic is injected by multi-qudit random untaries.

In the Haar case, all the spins in the bulk are fixed to $I$ in order to have lowest energy. The energy cost comes from the $S$ spins in the boundary region $A$ and the nearby bulk spins, giving
\begin{equation}
\langle M_{2n}\rangle=\log q\cdot|A| -S(A). \label{eq:RHC_SRE_mincut}
\end{equation}
It is half of mana in \eqref{eq: mana haar}.

\section{Clifford Average}
\label{appendix: clifford average}

The average over random Clifford unitaries produces a larger set of ``spins'' for the stat-mech model than the familiar one produced under Haar random unitaries. 
The technical result was developed in \cite{Gross_2021} and has been applied to entanglement calculation in \cite{Li:2021dbh}. In our paper, we use the technology to perform the calculation of quantum magic, more specifically, the Wigner function negativity and related magic measures, such as mana, in random Clifford circuits with and without measurements.  
We put the relevant notations and details here for the convenience of readers. 

Recall now that the allowed elements (``spins'') that survived under Clifford average 
\begin{equation}
    \int dU U^{\otimes t}FU^{\dagger\otimes t},
\end{equation}
where $F$ is an arbitrary operator. In the Haar case, it must commute with $V^{\otimes t}$ for an arbitrary unitary $V$. Due to Schur-Weyl duality, it can be decomposed as a linear combination of permutation operators. Now,  since we are averaging over Clifford unitaries instead, the moments only need to commute with tensor power of Clifford unitaries. The corresponding allowed ``spins'' (i.e. commutant of Clifford tensor power) are 
\begin{equation}
    r(T)=\sum_{(\boldx,\boldy)\in T}\ket{\boldx}\bra{\boldy},
\end{equation}
where $T$ is a stochastic Lagrangian subspace in $\mathbb{F}_q^{2t}$, defined by the following. 
\begin{definition}[stochastic Lagrangian subspace]
A subspace $T\subseteq \mathbb{F}_q^{2t}$ is a stochastic Lagrangian subspace if 
\begin{enumerate}
\item $\boldx\cdot\boldx=\boldy\cdot\boldy$ for all $(\boldx,\boldy)\in T$.
\item T has dimension $t$.
\item $\boldone\equiv (1,\cdots,1)\in T$.
\end{enumerate}
\end{definition}

The $r(T)$ operators are not full rank (hence not invertible) in general. The ones with full rank are associated with the elements of the stochastic orthogonal group $O_t(q)$, and can be expressed as follows, 
\begin{equation}
    r(O)=\sum_{\boldx\in T}\ket{O\boldx}\bra{\boldx}, \quad O\in O_t(q). 
\end{equation}
\begin{definition}[Stochastic orthogonal group]
The stochastic orthogonal group $O_t(q)$ consists of $t\times t$ matrices $O$ over $\mathbb{F}_q$ satisfying
\begin{itemize}
\item Orthogonal: $O\boldx\cdot O\boldx=\boldx\cdot \boldx$ for all $\boldx\in\mathbb{F}_q^t$ (in other words $O^TO=I$).
\item Stochastic: $O\boldone=\boldone$.
\end{itemize}
\end{definition}
It forms a group because the inverse is simply $O^{-1}=O^T$. We associate spaces and operators with the group elements as
\begin{equation}
    T_O\equiv\{(Ox,x)\},\qquad r(O)\equiv r(T_O)=\sum_x\ket{Ox}\bra{x}
\end{equation}

In parallel to the Haar case, one can define the coefficients of these spin elements inside the Clifford average as the generalized Weingarten functions~\cite{Li:2021dbh}
\begin{equation}
\Ens_V V^{\otimes t}\otimes V^{*\otimes t}=\sum_{T_a,T_b}\Wg_{q^2}(T_a,T_b)\big[r(T_a)\otimes I\ket{\tilde{I}_{q}}\bra{\tilde{I}_{q}}r(T_b)^{\dagger}\otimes I\big]^{\otimes 2}.
\label{eq:SM_Clifford_weingarten}
\end{equation}
Equivalently it can be written as
\begin{equation}
    \Ens_V V^{\otimes t}FV^{\dagger\otimes t}=\sum_{T_a,T_b}\Wg_{Q}(T_a,T_b)R(T_a)\Tr[R(T_b)^\dagger F],\quad Q=q^2,\quad R(T)=r(T)^{\otimes2}.\label{eq:SM_Clifford_weingarten2}
\end{equation}
It satisfies similar properties as the Haar case. For example, if we take $F=R(T_c)$ in \eqref{eq:SM_Clifford_weingarten2}, then using that $F$ commutes with any Clifford tensor powers, we get
\begin{equation}
    \sum_{T_a,T_b}\Wg_{Q}(T_a,T_b)R(T_a)Q^{t-|T_b,T_c|}=R(T_c),
\end{equation}
where $|T_b,T_c|$ is defined such that $\Tr[R(T_b)^\dagger R(T_c)]=Q^{t-|T_b,T_c|}$. It serves as a metric that take values in $\{0,1,\cdots,t\}$ and has the property that $|T_b,T_c|=0$ if an only if $T_b=T_c$\cite{Gross_2021,Li:2021dbh}. As shown in \cite{Gross_2021}, this set of basis is linear independent, so
\begin{equation}
    \sum_{T_b}\Wg_{Q}(T_a,T_b)Q^{t-|T_b,T_c|}=\delta_{T_a,T_c}. \label{eq:SM_Clifford_weingarten_inverse}
\end{equation}
Now that $|T_a,T_b|\ge 1$ for $T_a\neq T_c$, we can invert the matrix $Q^{t-|T_b,T_c|}$ in the large $Q$ limit to get
\begin{equation}
    \Wg_{Q}(T_a,T_b)=Q^{-t}\delta_{T_a,T_b}+\text{subleading}. \label{eq:SM_Clifford_weingarten_asymptotic}
\end{equation}
With \eqref{eq:SM_Clifford_weingarten_inverse} and \eqref{eq:SM_Clifford_weingarten_asymptotic}, the derivation of the triangular weights follows exactly from the Haar case, which yields
\begin{equation}
    \tilde{J}(T_a,T_a;T_b)=\delta_{T_a,T_b},\qquad \tilde{J}(T_a,T_b;T_a)=q^{-|T_a,T_b|}+\text{subleading}.
\end{equation}

\section{Details of the energy minimization}
\label{appendix: details of the energy minimization}

In the main text, the magic resource we inject into the systems are always a tensor product of local resource, i.e. the Haar unitaries are chosen independently for each qudit. An alternative way to inject magic is to consider a global Haar random unitary for the multi-qudit system.
\begin{equation}
\vcenter{\hbox{\includegraphics[width=0.34\textwidth]{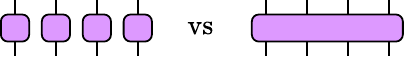}}}
\end{equation}
The difference between them is related to the total depth of non-Cliffordness in the circuit, known as the magic depth or T-depth~\cite{Amy_2013,tdepth,Zhang:2024jlz}.

From the stat-mech model perspective discussed in this paper, the difference is that for tensor products of single qudit magic injection, the boundary conditions can be freely chosen for each sites, while the global magic injection requires the boundary conditions to be the same for all the spins. 
In what follows we describe the detailed minimization procedure for magic injection from multi-qudit random unitaries, leading to a simple formula in the replica limit. In the single-qudit random unitary case, we will see that the replica limit is more involved.

In our circuit, suppose $A$ is the region whose mana is being computed, $M$ is the region where we inject magic by random unitaries or random measurements, and $B$ is the region being traced out with boundary condition $I$.

If the magic is injected by multi-qudit Haar random unitaries, then in $M$ the spins are either all $I$ or all $X$. We can minimize the energy within each of these possibilities, and finally minimize the contribution form the $I$ case and the $X$ case.
\begin{enumerate}
\item If all spins in $M$ are $I$, we can view $M$ as having the same boundary condition as $B$. We only need to account for the $(2n-1)l_{\bar{I},I}$ term in \eqref{eq:RCC_renyi_negativity_mincut}. In this case $l_{\bar{I},I}$ is the length of the cut separating the domain emanating from $A$ and the one from $B\cup M$, which we denote by $l_{A|BM}$. Minimizing the energy, we get $(2n-1)\min l_{A|BM}$.

\item If all spins in $M$ are $X$, we find it convenient to split the domain wall between $\bar{I}$ and $I$ into a domain wall between $\bar{I}$ and $X$ and a domain wall between $X$ and $I$. The splitting leaves the energy invariant because $X$ is along the geodesic between $\bar{I}$ and $I$\footnote{The distances satisfy $|\bar{I},X|+|X,I|=|\bar{I},I|$.}. For example
\begin{equation}
\vcenter{\hbox{\includegraphics[width=0.12\textwidth]{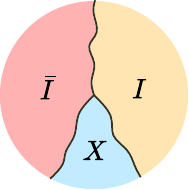}}}\quad
=\quad\vcenter{\hbox{\includegraphics[width=0.12\textwidth]{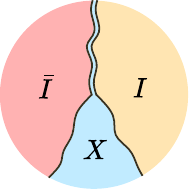}}}.
\end{equation}
Now $l_{\bar{I},X}$ can be viewed as the length of the cut separating the domain emanating from $A$ and another domain from $B\cup M$, which is $l_{A|BM}$ in our notation. Similarly, $l_{X,I}$ can be viewed as the length of the cut separating the domain emanating from $A\cup M$ and another domain from $B$, which we denote by $l_{AM|B}$. Now that we decoupled the $\bar{I}$ and $I$ domain with an intermediate domain $X$, we can minimize $l_{A|BM}$ and $l_{AM|B}$ independently to get less energetic configurations. The resulting contribution is $(n-1)\min l_{A|BM}+n\min l_{AM|B}$.
\end{enumerate}

Minimizing the two cases above, we get the Rényi negativity:
\begin{equation}
\begin{aligned}
    W^{(2n)}&=-\log q\min\big\{(2n-1)\min l_{A|BM},\ (n-1)\min l_{A|BM}+n\min l_{AM|B}\big\} \\
    &=-\log q\Big[(n-1)\min l_{A|BM}+n\min\big\{\min l_{A|BM},\ \min l_{AM|B}\big\}\Big].
    \label{eq:RCC_multiquditrandom_renyi_mana_mincut}
\end{aligned}
\end{equation}
Then the replica limit simply gives
\begin{equation}
\langle\calM\rangle=\max\big\{
    \min l_{A|BM}-\min l_{AM|B},\ 0\big\}\label{eq:RCC_multiquditrandom_mana_mincut}
\end{equation}

If magic is injected by single qudit random unitaries, the spins at different sites can be different. Let $M_X$ and $M_I$ denote the region in $M$ where the spins are $X$ and $I$ respectively. We should minimize over all possibilities of the bi-partition of $M$ into $M_X$ and $M_I$. For fixed $M_X$ and $M_I$, the minimization is similar to the second case above, if we replace $B\rightarrow B\cup M_I$ and $M\rightarrow M_X$. The final result is
\begin{equation}
\begin{aligned}
    W^{(2n)}&=-\log q\min_{M_X,M_I}\big[(n-1)\min l_{A|BM}+n\min l_{ABM_I|M_X}\big]
\end{aligned}
\end{equation}
It is more difficult to take the replica limit here, because of the $\min_{M_X,M_I}$ procedure.

\bibliography{reference.bib}

\providecommand{\href}[2]{#2}\begingroup\raggedright\begin{thebibliography}{10}

\bibitem{RevModPhys.80.517}
L.~Amico, R.~Fazio, A.~Osterloh and V.~Vedral, \emph{Entanglement in many-body systems}, \href{http://dx.doi.org/10.1103/RevModPhys.80.517}{\emph{Rev. Mod. Phys.} {\bf 80} (May, 2008) 517--576}.

\bibitem{RevModPhys.81.865}
R.~Horodecki, P.~Horodecki, M.~Horodecki and K.~Horodecki, \emph{Quantum entanglement}, \href{http://dx.doi.org/10.1103/RevModPhys.81.865}{\emph{Rev. Mod. Phys.} {\bf 81} (Jun, 2009) 865--942}.

\bibitem{Gottesman:1998hu}
D.~Gottesman, \emph{{The Heisenberg representation of quantum computers}},  in \emph{{22nd International Colloquium on Group Theoretical Methods in Physics}}, pp.~32--43, 7, 1998.
\newblock \href{https://arxiv.org/abs/quant-ph/9807006}{{\tt quant-ph/9807006}}.

\bibitem{Calderbank:1996hm}
A.~R. Calderbank, E.~M. Rains, P.~W. Shor and N.~J.~A. Sloane, \emph{{Quantum error correction and orthogonal geometry}}, \href{http://dx.doi.org/10.1103/PhysRevLett.78.405}{\emph{Phys. Rev. Lett.} {\bf 78} (1997) 405--408}, [\href{https://arxiv.org/abs/quant-ph/9605005}{{\tt quant-ph/9605005}}].

\bibitem{Gottesman:1997zz}
D.~Gottesman, \emph{{Stabilizer codes and quantum error correction}},  \href{https://arxiv.org/abs/quant-ph/9705052}{{\tt quant-ph/9705052}}.

\bibitem{PhysRevA.71.022316}
S.~Bravyi and A.~Kitaev, \emph{Universal quantum computation with ideal clifford gates and noisy ancillas}, \href{http://dx.doi.org/10.1103/PhysRevA.71.022316}{\emph{Phys. Rev. A} {\bf 71} (Feb, 2005) 022316}.

\bibitem{knill05}
E.~Knill, \emph{Quantum computing with realistically noisy devices}, \href{http://dx.doi.org/10.1038/nature03350}{\emph{Nature} {\bf 434} (2005) 39--44}.

\bibitem{Veitch_2012}
V.~Veitch, C.~Ferrie, D.~Gross and J.~Emerson, \emph{Negative quasi-probability as a resource for quantum computation}, \href{http://dx.doi.org/10.1088/1367-2630/14/11/113011}{\emph{New Journal of Physics} {\bf 14} (Nov., 2012) 113011}.

\bibitem{Emerson:2013zse}
J.~Emerson, D.~Gottesman, S.~A.~H. Mousavian and V.~Veitch, \emph{{The resource theory of stabilizer quantum computation}}, \href{http://dx.doi.org/10.1088/1367-2630/16/1/013009}{\emph{New J. Phys.} {\bf 16} (2014) 013009}, [\href{https://arxiv.org/abs/1307.7171}{{\tt 1307.7171}}].

\bibitem{PhysRevLett.118.090501}
M.~Howard and E.~Campbell, \emph{Application of a resource theory for magic states to fault-tolerant quantum computing}, \href{http://dx.doi.org/10.1103/PhysRevLett.118.090501}{\emph{Phys. Rev. Lett.} {\bf 118} (Mar, 2017) 090501}.

\bibitem{PhysRevA.97.062332}
M.~Ahmadi, H.~B. Dang, G.~Gour and B.~C. Sanders, \emph{Quantification and manipulation of magic states}, \href{http://dx.doi.org/10.1103/PhysRevA.97.062332}{\emph{Phys. Rev. A} {\bf 97} (Jun, 2018) 062332}.

\bibitem{seddon2019quantifying}
J.~R. Seddon and E.~T. Campbell, \emph{Quantifying magic for multi-qubit operations}, {\emph{Proceedings of the Royal Society A} {\bf 475} (2019) 20190251}.

\bibitem{Wang_2019}
X.~Wang, M.~M. Wilde and Y.~Su, \emph{Quantifying the magic of quantum channels}, \href{http://dx.doi.org/10.1088/1367-2630/ab451d}{\emph{New Journal of Physics} {\bf 21} (oct, 2019) 103002}.

\bibitem{Liu:2020yso}
Z.-W. Liu and A.~Winter, \emph{{Many-Body Quantum Magic}}, \href{http://dx.doi.org/10.1103/PRXQuantum.3.020333}{\emph{PRX Quantum} {\bf 3} (2022) 020333}, [\href{https://arxiv.org/abs/2010.13817}{{\tt 2010.13817}}].

\bibitem{Rattacaso:2023kzm}
D.~Rattacaso, L.~Leone, S.~F.~E. Oliviero and A.~Hamma, \emph{{Stabilizer entropy dynamics after a quantum quench}}, \href{http://dx.doi.org/10.1103/PhysRevA.108.042407}{\emph{Phys. Rev. A} {\bf 108} (2023) 042407}, [\href{https://arxiv.org/abs/2304.13768}{{\tt 2304.13768}}].

\bibitem{Lopez:2024jjq}
J.~A. M.~n. L\'opez and P.~Kos, \emph{{Exact solution of long-range stabilizer R\'enyi entropy in the dual-unitary XXZ model}},  \href{https://arxiv.org/abs/2405.04448}{{\tt 2405.04448}}.

\bibitem{Passarelli:2024lpm}
G.~Passarelli, P.~Lucignano, D.~Rossini and A.~Russomanno, \emph{{Chaos and magic in the dissipative quantum kicked top}},  \href{https://arxiv.org/abs/2406.16585}{{\tt 2406.16585}}.

\bibitem{Leone:2023uqw}
L.~Leone, S.~F.~E. Oliviero, G.~Esposito and A.~Hamma, \emph{{Phase transition in stabilizer entropy and efficient purity estimation}}, \href{http://dx.doi.org/10.1103/PhysRevA.109.032403}{\emph{Phys. Rev. A} {\bf 109} (2024) 032403}, [\href{https://arxiv.org/abs/2302.07895}{{\tt 2302.07895}}].

\bibitem{Niroula:2023meg}
P.~Niroula, C.~D. White, Q.~Wang, S.~Johri, D.~Zhu, C.~Monroe et~al., \emph{{Phase transition in magic with random quantum circuits}},  \href{https://arxiv.org/abs/2304.10481}{{\tt 2304.10481}}.

\bibitem{Tarabunga:2024din}
P.~S. Tarabunga and E.~Tirrito, \emph{{Magic transition in measurement-only circuits}},  \href{https://arxiv.org/abs/2407.15939}{{\tt 2407.15939}}.

\bibitem{Bejan:2023zqm}
M.~Bejan, C.~McLauchlan and B.~B\'eri, \emph{{Dynamical Magic Transitions in Monitored Clifford+T Circuits}}, \href{http://dx.doi.org/10.1103/PRXQuantum.5.030332}{\emph{PRX Quantum} {\bf 5} (2024) 030332}, [\href{https://arxiv.org/abs/2312.00132}{{\tt 2312.00132}}].

\bibitem{Fux:2023brx}
G.~E. Fux, E.~Tirrito, M.~Dalmonte and R.~Fazio, \emph{{Entanglement-magic separation in hybrid quantum circuits}},  \href{https://arxiv.org/abs/2312.02039}{{\tt 2312.02039}}.

\bibitem{Wang:2023uog}
Y.~Wang and Y.~Li, \emph{{Stabilizer R\'enyi entropy on qudits}}, \href{http://dx.doi.org/10.1007/s11128-023-04186-9}{\emph{Quant. Inf. Proc.} {\bf 22} (2023) 444}.

\bibitem{Zhou_2020}
S.~Zhou, Z.~Yang, A.~Hamma and C.~Chamon, \emph{Single t gate in a clifford circuit drives transition to universal entanglement spectrum statistics}, \href{http://dx.doi.org/10.21468/scipostphys.9.6.087}{\emph{SciPost Physics} {\bf 9} (Dec., 2020) }.

\bibitem{Turkeshi:2024pnj}
X.~Turkeshi, E.~Tirrito and P.~Sierant, \emph{{Magic spreading in random quantum circuits}},  \href{https://arxiv.org/abs/2407.03929}{{\tt 2407.03929}}.

\bibitem{Chen:2022yza}
L.~Chen, R.~J. Garcia, K.~Bu and A.~Jaffe, \emph{{Magic of random matrix product states}}, \href{http://dx.doi.org/10.1103/PhysRevB.109.174207}{\emph{Phys. Rev. B} {\bf 109} (2024) 174207}, [\href{https://arxiv.org/abs/2211.10350}{{\tt 2211.10350}}].

\bibitem{deutsch1991quantum}
J.~M. Deutsch, \emph{Quantum statistical mechanics in a closed system}, {\emph{Physical Review A} {\bf 43} (1991) 2046--2049}.

\bibitem{srednicki1994chaos}
M.~Srednicki, \emph{Chaos and quantum thermalization}, {\emph{Physical Review E} {\bf 50} (1994) 888--901}.

\bibitem{bekenstein1973black}
J.~D. Bekenstein, \emph{Black holes and entropy}, {\emph{Physical Review D} {\bf 7} (1973) 2333--2346}.

\bibitem{hawking1975particle}
S.~W. Hawking, \emph{Particle creation by black holes}, {\emph{Communications in Mathematical Physics} {\bf 43} (1975) 199--220}.

\bibitem{ryu2006holographic}
S.~Ryu and T.~Takayanagi, \emph{Holographic derivation of entanglement entropy from the anti--de sitter space/conformal field theory correspondence}, {\emph{Physical Review Letters} {\bf 96} (2006) 181602}.

\bibitem{PhysRevLett.115.070501}
H.~Pashayan, J.~J. Wallman and S.~D. Bartlett, \emph{Estimating outcome probabilities of quantum circuits using quasiprobabilities}, \href{http://dx.doi.org/10.1103/PhysRevLett.115.070501}{\emph{Phys. Rev. Lett.} {\bf 115} (Aug, 2015) 070501}.

\bibitem{Fisher:2022qey}
M.~P.~A. Fisher, V.~Khemani, A.~Nahum and S.~Vijay, \emph{{Random Quantum Circuits}}, \href{http://dx.doi.org/10.1146/annurev-conmatphys-031720-030658}{\emph{Ann. Rev. Condensed Matter Phys.} {\bf 14} (2023) 335--379}, [\href{https://arxiv.org/abs/2207.14280}{{\tt 2207.14280}}].

\bibitem{Haug:2023hcs}
T.~Haug and L.~Piroli, \emph{{Stabilizer entropies and nonstabilizerness monotones}}, \href{http://dx.doi.org/10.22331/q-2023-08-28-1092}{\emph{Quantum} {\bf 7} (2023) 1092}, [\href{https://arxiv.org/abs/2303.10152}{{\tt 2303.10152}}].

\bibitem{Leone:2024lfr}
L.~Leone and L.~Bittel, \emph{{Stabilizer entropies are monotones for magic-state resource theory}},  \href{https://arxiv.org/abs/2404.11652}{{\tt 2404.11652}}.

\bibitem{Gross_2006}
D.~Gross, \emph{Hudson’s theorem for finite-dimensional quantum systems}, \href{http://dx.doi.org/10.1063/1.2393152}{\emph{Journal of Mathematical Physics} {\bf 47} (Dec., 2006) }.

\bibitem{Delfosse_2017}
N.~Delfosse, C.~Okay, J.~Bermejo-Vega, D.~E. Browne and R.~Raussendorf, \emph{Equivalence between contextuality and negativity of the wigner function for qudits}, \href{http://dx.doi.org/10.1088/1367-2630/aa8fe3}{\emph{New Journal of Physics} {\bf 19} (Dec., 2017) 123024}.

\bibitem{Pashayan_2015}
H.~Pashayan, J.~J. Wallman and S.~D. Bartlett, \emph{Estimating outcome probabilities of quantum circuits using quasiprobabilities}, \href{http://dx.doi.org/10.1103/physrevlett.115.070501}{\emph{Physical Review Letters} {\bf 115} (Aug., 2015) }.

\bibitem{Gross_2021}
D.~Gross, S.~Nezami and M.~Walter, \emph{Schur–weyl duality for the clifford group with applications: Property testing, a robust hudson theorem, and de finetti representations}, \href{http://dx.doi.org/10.1007/s00220-021-04118-7}{\emph{Communications in Mathematical Physics} {\bf 385} (June, 2021) 1325–1393}.

\bibitem{White2020ManaIH}
C.~D. White and J.~H. Wilson, \emph{Mana in haar-random states}, {\emph{arXiv preprint arXiv:2011.13937} (2020) }.

\bibitem{White:2020zoz}
C.~D. White, C.~Cao and B.~Swingle, \emph{{Conformal field theories are magical}}, \href{http://dx.doi.org/10.1103/PhysRevB.103.075145}{\emph{Phys. Rev. B} {\bf 103} (2021) 075145}, [\href{https://arxiv.org/abs/2007.01303}{{\tt 2007.01303}}].

\bibitem{Goto:2021anl}
K.~Goto, T.~Nosaka and M.~Nozaki, \emph{{Probing chaos by magic monotones}}, \href{http://dx.doi.org/10.1103/PhysRevD.106.126009}{\emph{Phys. Rev. D} {\bf 106} (2022) 126009}, [\href{https://arxiv.org/abs/2112.14593}{{\tt 2112.14593}}].

\bibitem{Sewell:2022lao}
T.~J. Sewell and C.~D. White, \emph{{Mana and thermalization: Probing the feasibility of near-Clifford Hamiltonian simulation}}, \href{http://dx.doi.org/10.1103/PhysRevB.106.125130}{\emph{Phys. Rev. B} {\bf 106} (2022) 125130}, [\href{https://arxiv.org/abs/2201.12367}{{\tt 2201.12367}}].

\bibitem{Ahmadi:2022bkg}
A.~Ahmadi and E.~Greplova, \emph{{Quantifying non-stabilizerness via information scrambling}}, \href{http://dx.doi.org/10.21468/SciPostPhys.16.2.043}{\emph{SciPost Phys.} {\bf 16} (2024) 043}, [\href{https://arxiv.org/abs/2204.11236}{{\tt 2204.11236}}].

\bibitem{Tarabunga:2023hau}
P.~S. Tarabunga, \emph{{Critical behaviors of non-stabilizerness in quantum spin chains}}, \href{http://dx.doi.org/10.22331/q-2024-07-17-1413}{\emph{Quantum} {\bf 8} (2024) 1413}, [\href{https://arxiv.org/abs/2309.00676}{{\tt 2309.00676}}].

\bibitem{Bravyi_2016im}
S.~Bravyi and D.~Gosset, \emph{Improved classical simulation of quantum circuits dominated by clifford gates}, \href{http://dx.doi.org/10.1103/physrevlett.116.250501}{\emph{Physical Review Letters} {\bf 116} (June, 2016) }.

\bibitem{Bravyi_2016tr}
S.~Bravyi, G.~Smith and J.~A. Smolin, \emph{Trading classical and quantum computational resources}, \href{http://dx.doi.org/10.1103/physrevx.6.021043}{\emph{Physical Review X} {\bf 6} (June, 2016) }.

\bibitem{Bravyi_2019}
S.~Bravyi, D.~Browne, P.~Calpin, E.~Campbell, D.~Gosset and M.~Howard, \emph{Simulation of quantum circuits by low-rank stabilizer decompositions}, \href{http://dx.doi.org/10.22331/q-2019-09-02-181}{\emph{Quantum} {\bf 3} (Sept., 2019) 181}.

\bibitem{Seddon_2021}
J.~R. Seddon, B.~Regula, H.~Pashayan, Y.~Ouyang and E.~T. Campbell, \emph{Quantifying quantum speedups: Improved classical simulation from tighter magic monotones}, \href{http://dx.doi.org/10.1103/prxquantum.2.010345}{\emph{PRX Quantum} {\bf 2} (Mar., 2021) }.

\bibitem{Howard_2017}
M.~Howard and E.~Campbell, \emph{Application of a resource theory for magic states to fault-tolerant quantum computing}, \href{http://dx.doi.org/10.1103/physrevlett.118.090501}{\emph{Physical Review Letters} {\bf 118} (Mar., 2017) }.

\bibitem{Beverland:2019jej}
M.~Beverland, E.~Campbell, M.~Howard and V.~Kliuchnikov, \emph{{Lower bounds on the non-Clifford resources for quantum computations}}, \href{http://dx.doi.org/10.1088/2058-9565/ab8963}{\emph{Quantum Sci. Technol.} {\bf 5} (2020) 035009}, [\href{https://arxiv.org/abs/1904.01124}{{\tt 1904.01124}}].

\bibitem{Wang_2020}
X.~Wang, M.~M. Wilde and Y.~Su, \emph{Efficiently computable bounds for magic state distillation}, \href{http://dx.doi.org/10.1103/physrevlett.124.090505}{\emph{Physical Review Letters} {\bf 124} (Mar., 2020) }.

\bibitem{Leone:2021rzd}
L.~Leone, S.~F.~E. Oliviero and A.~Hamma, \emph{{Stabilizer R\'enyi Entropy}}, \href{http://dx.doi.org/10.1103/PhysRevLett.128.050402}{\emph{Phys. Rev. Lett.} {\bf 128} (2022) 050402}, [\href{https://arxiv.org/abs/2106.12587}{{\tt 2106.12587}}].

\bibitem{Nahum:2017yvy}
A.~Nahum, S.~Vijay and J.~Haah, \emph{{Operator Spreading in Random Unitary Circuits}}, \href{http://dx.doi.org/10.1103/PhysRevX.8.021014}{\emph{Phys. Rev. X} {\bf 8} (2018) 021014}, [\href{https://arxiv.org/abs/1705.08975}{{\tt 1705.08975}}].

\bibitem{Zhou:2018myl}
T.~Zhou and A.~Nahum, \emph{{Emergent statistical mechanics of entanglement in random unitary circuits}}, \href{http://dx.doi.org/10.1103/PhysRevB.99.174205}{\emph{Phys. Rev. B} {\bf 99} (2019) 174205}, [\href{https://arxiv.org/abs/1804.09737}{{\tt 1804.09737}}].

\bibitem{Hunter-Jones:2019lps}
N.~Hunter-Jones, \emph{{Unitary designs from statistical mechanics in random quantum circuits}},  \href{https://arxiv.org/abs/1905.12053}{{\tt 1905.12053}}.

\bibitem{Bao:2019qah}
Y.~Bao, S.~Choi and E.~Altman, \emph{{Theory of the phase transition in random unitary circuits with measurements}}, \href{http://dx.doi.org/10.1103/PhysRevB.101.104301}{\emph{Phys. Rev. B} {\bf 101} (2020) 104301}, [\href{https://arxiv.org/abs/1908.04305}{{\tt 1908.04305}}].

\bibitem{Jian:2019mny}
C.-M. Jian, Y.-Z. You, R.~Vasseur and A.~W.~W. Ludwig, \emph{{Measurement-induced criticality in random quantum circuits}}, \href{http://dx.doi.org/10.1103/PhysRevB.101.104302}{\emph{Phys. Rev. B} {\bf 101} (2020) 104302}, [\href{https://arxiv.org/abs/1908.08051}{{\tt 1908.08051}}].

\bibitem{Collins_2003}
B.~Collins, \emph{{Moments and cumulants of polynomial random variables on unitarygroups, the Itzykson-Zuber integral, and free probability}}, \href{http://dx.doi.org/10.1155/S107379280320917X}{\emph{International Mathematics Research Notices} {\bf 2003} (01, 2003) 953--982}, [\href{https://arxiv.org/abs/https://academic.oup.com/imrn/article-pdf/2003/17/953/1881428/2003-17-953.pdf}{{\tt https://academic.oup.com/imrn/article-pdf/2003/17/953/1881428/2003-17-953.pdf}}].

\bibitem{Collins_2006}
B.~Collins and P.~Śniady, \emph{Integration with respect to the haar measure on unitary, orthogonal and symplectic group}, \href{http://dx.doi.org/10.1007/s00220-006-1554-3}{\emph{Communications in Mathematical Physics} {\bf 264} (Mar., 2006) 773–795}.

\bibitem{Raussendorf:2001zim}
R.~Raussendorf and H.~J. Briegel, \emph{{A One-Way Quantum Computer}}, \href{http://dx.doi.org/10.1103/PhysRevLett.86.5188}{\emph{Phys. Rev. Lett.} {\bf 86} (2001) 5188}.

\bibitem{Miller_2016}
J.~Miller and A.~Miyake, \emph{Hierarchy of universal entanglement in 2d measurement-based quantum computation}, \href{http://dx.doi.org/10.1038/npjqi.2016.36}{\emph{npj Quantum Information} {\bf 2} (Nov., 2016) }.

\bibitem{Li:2021dbh}
Y.~Li, R.~Vasseur, M.~P.~A. Fisher and A.~W.~W. Ludwig, \emph{{Statistical mechanics model for Clifford random tensor networks and monitored quantum circuits}}, \href{http://dx.doi.org/10.1103/PhysRevB.109.174307}{\emph{Phys. Rev. B} {\bf 109} (2024) 174307}, [\href{https://arxiv.org/abs/2110.02988}{{\tt 2110.02988}}].

\bibitem{Zhu_2017}
H.~Zhu, \emph{Multiqubit clifford groups are unitary 3-designs}, \href{http://dx.doi.org/10.1103/physreva.96.062336}{\emph{Physical Review A} {\bf 96} (Dec., 2017) }.

\bibitem{Webb:2015vwz}
Z.~Webb, \emph{{The Clifford group forms a unitary 3-design}}, \href{http://dx.doi.org/10.26421/QIC16.15-16-8}{\emph{Quant. Inf. Comput.} {\bf 16} (2016) 1379--1400}, [\href{https://arxiv.org/abs/1510.02769}{{\tt 1510.02769}}].

\bibitem{Schumacher:1996dy}
B.~Schumacher and M.~A. Nielsen, \emph{{Quantum data processing and error correction}}, \href{http://dx.doi.org/10.1103/PhysRevA.54.2629}{\emph{Phys. Rev. A} {\bf 54} (1996) 2629}, [\href{https://arxiv.org/abs/quant-ph/9604022}{{\tt quant-ph/9604022}}].

\bibitem{Lloyd:1996at}
S.~Lloyd, \emph{{The Capacity of the noisy quantum channel}}, \href{http://dx.doi.org/10.1103/PhysRevA.55.1613}{\emph{Phys. Rev. A} {\bf 55} (1997) 1613}, [\href{https://arxiv.org/abs/quant-ph/9604015}{{\tt quant-ph/9604015}}].

\bibitem{Gullans:2020whl}
M.~J. Gullans, S.~Krastanov, D.~A. Huse, L.~Jiang and S.~T. Flammia, \emph{{Quantum Coding with Low-Depth Random Circuits}}, \href{http://dx.doi.org/10.1103/PhysRevX.11.031066}{\emph{Phys. Rev. X} {\bf 11} (2021) 031066}, [\href{https://arxiv.org/abs/2010.09775}{{\tt 2010.09775}}].

\bibitem{Lami:2024osd}
G.~Lami, T.~Haug and J.~De~Nardis, \emph{{Quantum State Designs with Clifford Enhanced Matrix Product States}},  \href{https://arxiv.org/abs/2404.18751}{{\tt 2404.18751}}.

\bibitem{Qian:2024vea}
X.~Qian, J.~Huang and M.~Qin, \emph{{Augmenting Density Matrix Renormalization Group with Clifford Circuits}},  \href{https://arxiv.org/abs/2405.09217}{{\tt 2405.09217}}.

\bibitem{Qian:2024wun}
X.~Qian, J.~Huang and M.~Qin, \emph{{Clifford Circuits Augmented Time-Dependent Variational Principle}},  \href{https://arxiv.org/abs/2407.03202}{{\tt 2407.03202}}.

\bibitem{Mello:2024jev}
A.~F. Mello, A.~Santini, G.~Lami, J.~De~Nardis and M.~Collura, \emph{{Clifford Dressed Time-Dependent Variational Principle}},  \href{https://arxiv.org/abs/2407.01692}{{\tt 2407.01692}}.

\bibitem{Qian:2024lqx}
X.~Qian, J.~Huang and M.~Qin, \emph{{Augmenting Finite Temperature Tensor Network with Clifford Circuits}},  \href{https://arxiv.org/abs/2410.15709}{{\tt 2410.15709}}.

\bibitem{True:2022ypf}
S.~True and A.~Hamma, \emph{{Transitions in Entanglement Complexity in Random Circuits}}, \href{http://dx.doi.org/10.22331/q-2022-09-22-818}{\emph{Quantum} {\bf 6} (2022) 818}, [\href{https://arxiv.org/abs/2202.02648}{{\tt 2202.02648}}].

\bibitem{Tirrito:2023fnw}
E.~Tirrito, P.~S. Tarabunga, G.~Lami, T.~Chanda, L.~Leone, S.~F.~E. Oliviero et~al., \emph{{Quantifying nonstabilizerness through entanglement spectrum flatness}}, \href{http://dx.doi.org/10.1103/PhysRevA.109.L040401}{\emph{Phys. Rev. A} {\bf 109} (2024) L040401}, [\href{https://arxiv.org/abs/2304.01175}{{\tt 2304.01175}}].

\bibitem{Cao:2024nrx}
C.~Cao, G.~Cheng, A.~Hamma, L.~Leone, W.~Munizzi and S.~F.~E. Oliviero, \emph{{Gravitational back-reaction is magical}},  \href{https://arxiv.org/abs/2403.07056}{{\tt 2403.07056}}.

\bibitem{Gu:2024qvn}
A.~Gu, S.~F.~E. Oliviero and L.~Leone, \emph{{Magic-induced computational separation in entanglement theory}},  \href{https://arxiv.org/abs/2403.19610}{{\tt 2403.19610}}.

\bibitem{Li:2021par}
Y.~Li, S.~Vijay and M.~P.~A. Fisher, \emph{{Entanglement Domain Walls in Monitored Quantum Circuits and the Directed Polymer in a Random Environment}}, \href{http://dx.doi.org/10.1103/PRXQuantum.4.010331}{\emph{PRX Quantum} {\bf 4} (2023) 010331}, [\href{https://arxiv.org/abs/2105.13352}{{\tt 2105.13352}}].

\bibitem{Li:2022ycx}
Z.~Li, S.~Sang and T.~H. Hsieh, \emph{{Entanglement dynamics of noisy random circuits}}, \href{http://dx.doi.org/10.1103/PhysRevB.107.014307}{\emph{Phys. Rev. B} {\bf 107} (2023) 014307}, [\href{https://arxiv.org/abs/2203.16555}{{\tt 2203.16555}}].

\bibitem{Dias:2022goz}
B.~C. Dias, D.~Perkovic, M.~Haque, P.~Ribeiro and P.~A. McClarty, \emph{{Quantum noise as a symmetry-breaking field}}, \href{http://dx.doi.org/10.1103/PhysRevB.108.L060302}{\emph{Phys. Rev. B} {\bf 108} (2023) L060302}, [\href{https://arxiv.org/abs/2208.13861}{{\tt 2208.13861}}].

\bibitem{Li:2023ekc}
Y.~Li and M.~Claassen, \emph{{Statistical mechanics of monitored dissipative random circuits}}, \href{http://dx.doi.org/10.1103/PhysRevB.108.104310}{\emph{Phys. Rev. B} {\bf 108} (2023) 104310}, [\href{https://arxiv.org/abs/2303.08152}{{\tt 2303.08152}}].

\bibitem{Liu:2024pyz}
S.~Liu, M.-R. Li, S.-X. Zhang and S.-K. Jian, \emph{{Entanglement structure and information protection in noisy hybrid quantum circuits}},  \href{https://arxiv.org/abs/2401.01593}{{\tt 2401.01593}}.

\bibitem{Leone_2021}
L.~Leone, S.~F.~E. Oliviero, Y.~Zhou and A.~Hamma, \emph{Quantum chaos is quantum}, \href{http://dx.doi.org/10.22331/q-2021-05-04-453}{\emph{Quantum} {\bf 5} (May, 2021) 453}.

\bibitem{PhysRevLett.132.240602}
T.~Haug, S.~Lee and M.~S. Kim, \emph{Efficient quantum algorithms for stabilizer entropies}, \href{http://dx.doi.org/10.1103/PhysRevLett.132.240602}{\emph{Phys. Rev. Lett.} {\bf 132} (Jun, 2024) 240602}.

\bibitem{Campbell_2011}
E.~T. Campbell, \emph{Catalysis and activation of magic states in fault-tolerant architectures}, \href{http://dx.doi.org/10.1103/physreva.83.032317}{\emph{Physical Review A} {\bf 83} (Mar., 2011) }.

\bibitem{Bu:2023ssg}
K.~Bu, W.~Gu and A.~Jaffe, \emph{{Quantum Entropy and Central Limit Theorem}}, \href{http://dx.doi.org/10.1073/pnas.2304589120}{\emph{Proc. Nat. Acad. Sci.} {\bf 120} (2023) e2304589120}, [\href{https://arxiv.org/abs/2302.07841}{{\tt 2302.07841}}].

\bibitem{Bu:2023uic}
K.~Bu, W.~Gu and A.~Jaffe, \emph{{Discrete Quantum Gaussians and Central Limit Theorem}},  \href{https://arxiv.org/abs/2302.08423}{{\tt 2302.08423}}.

\bibitem{Turkeshi:2023lqu}
X.~Turkeshi, A.~Dymarsky and P.~Sierant, \emph{{Pauli Spectrum and Magic of Typical Quantum Many-Body States}},  \href{https://arxiv.org/abs/2312.11631}{{\tt 2312.11631}}.

\bibitem{Amy_2013}
M.~Amy, D.~Maslov, M.~Mosca and M.~Roetteler, \emph{A meet-in-the-middle algorithm for fast synthesis of depth-optimal quantum circuits}, \href{http://dx.doi.org/10.1109/tcad.2013.2244643}{\emph{IEEE Transactions on Computer-Aided Design of Integrated Circuits and Systems} {\bf 32} (June, 2013) 818–830}.

\bibitem{tdepth}
P.~Selinger, \emph{Quantum circuits of $t$-depth one}, \href{http://dx.doi.org/10.1103/PhysRevA.87.042302}{\emph{Phys. Rev. A} {\bf 87} (Apr, 2013) 042302}.

\bibitem{Zhang:2024jlz}
Y.~Zhang and Y.~Zhang, \emph{{Classical Simulability of Quantum Circuits with Shallow Magic Depth}},  \href{https://arxiv.org/abs/2409.13809}{{\tt 2409.13809}}.

\end{thebibliography}\endgroup

\end{document}